\documentclass{article}

\usepackage{graphicx}
\usepackage{amsmath}
\usepackage{amssymb}
\usepackage{mathtools}
\usepackage{setspace}
\usepackage{subcaption}
\usepackage[hidelinks]{hyperref}

\usepackage{lineno}
\usepackage{sansmath} 
\sansmath

\usepackage{grffile}
\usepackage[a4paper, portrait, margin=2cm]{geometry}
\usepackage{color}

\graphicspath{{Kostas/Output/}}

\title{Encounter networks from collective mitochondrial dynamics support the emergence of effective mtDNA genomes in plant cells \\ \normalsize \vspace{0.5cm} Konstantinos Giannakis${}^{1}$, Joanna M. Chustecki${}^{2}$, Iain G. Johnston${}^{1,3,*}$\\
${}^{1}$ Department of Mathematics, University of Bergen, Bergen, Norway \\
  ${}^{2}$ School of Biosciences, University of Birmingham, Birmingham, UK \\
  ${}^{3}$ Computational Biology Unit, University of Bergen, Bergen, Norway \\
  ${}^{*}$ correspondence to \url{iain.johnston@uib.no}
}
\date{}
\begin{document}
\maketitle

\section*{Abstract}
Mitochondria in plant cells form strikingly dynamic populations of largely individual organelles. Each mitochondrion contains on average less than a full copy of the mitochondrial DNA (mtDNA) genome. Here, we asked whether mitochondrial dynamics may allow individual mitochondria to `collect' a full copy of the mtDNA genome over time, by facilitating exchange between individuals. Akin to trade on a social network, exchange of mtDNA fragments across organelles may lead to the emergence of full `effective' genomes in individuals over time. We characterise the collective dynamics of mitochondria in \emph{Arabidopsis thaliana} hypocotyl cells using a recent approach combining single-cell timelapse microscopy, video analysis, and network science. We then use a quantitative model to predict the capacity for the sharing and accumulation of genetic information through the networks of encounters between mitochondria. We find that biological encounter networks are strikingly well predisposed to support the collection of full genomes over time, outperforming a range of other networks generated from theory and simulation. Using results from the coupon collector's problem, we show that the upper tail of the degree distribution is a key determinant of an encounter network's performance at this task and discuss how features of mitochondrial dynamics observed in biology facilitate the emergence of full effective genomes.

\section*{Introduction}
Mitochondria are vital bioenergetic organelles, present in the vast majority of eukaryotic cells. Across and within eukaryotic organisms, mitochondria display a diverse variety of forms and dynamics. In plant cells, mitochondria largely exist as discrete, independent organelles. Unlike metazoan and fungal mitochondria, they rarely form large physical networks (with some exceptions \cite{seguisimarro2009mitochondrial}). Individual plant mitochondria are highly dynamic, moving rapidly through the cell both along the cytoskeleton and diffusively \cite{logan2000mitochondria, logan2006mitochondrial}.

This physical population has a coupled genetic structure. Plant mitochondria do not typically contain full copies of the mtDNA genome \cite{preuten2010fewer, takanashi2006different, johnston2019tension}. Instead, many mitochondria either contain mtDNA `subgenomic' molecules -- encoding a reduced subset of mtDNA genes -- or no mtDNA at all. The question arises: how do plant mitochondria maintain their protein complements, without a complete local genome from which to express new proteins?

One possibility \cite{arimura2004frequent, logan2006mitochondrial, takanashi2006different, arimura2018fission} is that exchanges of mtDNA subsets between individuals can, over time, lead to the emergence of full `effective' mtDNA genomes in individuals over time. For example, picture a genome which can be partitioned into two regions, A and B. One mitochondrion initially possesses a subgenomic molecule containing only region A of the genome. Another initially possesses only region B. Each expresses the genes contained in its subgenomic region. Then the two mitochondria physically meet and exchange their subgenomic molecules. The first mitochondrion can now express genes from B, and vice versa. Indeed, within the dynamic cellular population of mitochondria, transient colocalisations occur, resembling `kiss-and-run' events in bacterial populations \cite{liu2009mitochondrial, logan2010mitochondrial,el2014friendly,chustecki2021network}. Some of these colocalisations result in transient fusion between two mitochondria. When fusion occurs, mitochondria can exchange genetic and protein material: indeed, mixing occurs through the entire cellular population on a timescale of hours \cite{arimura2004frequent}.

Recent work has characterised the `encounter networks' between mitochondria in plant cells, describing which mitochondria encounter which others over time \cite{chustecki2021network}. Here, mitochondria are nodes, with two nodes linked by an edge if the corresponding mitochondria have been recorded within a threshold distance. Chustecki \emph{et al.} showed that these encounter networks have structures which have the potential to facilitate efficient exchange of content, while also allowing mitochondria to spread evenly through the cell \cite{chustecki2021network}. Hence, mitochondrial dynamics have the potential to resolve a tension between competing cell priorities: even spacing of mitochondria (with metabolic and energetic advantages) and colocalisation of mitochondria (for beneficial exchange of contents). These principles support the developing cell biological perspective of inter-organelle interactions \cite{valm2017applying, cohen2018interacting, picard2021social}.

Such functional encounters are an example of emergence, where the behaviour of a collective of individuals is different from the sum of individual behaviours. There are two coupled instances of emergent behaviour in our system -- physical and genetic. First, the encounter network of mitochondria emerges from their underlying physical dynamics in the cell \cite{williams2019connect}. Second, through genetic exchanges within this encounter network, an `effective genome' for each mitochondrion may emerge. That is, over time, each mitochondrion will be exposed to a growing set of genetic information. We hypothesised that the exchange efficiency of encounter networks could allow a mechanism for plant mitochondria to address their maintenance problem. Specifically, if mitochondria can efficiently exchange genetic information, then the effective genome, to which each mitochondrion is exposed over time, may eventually grow to include the full set of genes in the full genome. To investigate this hypothesis, we proceed by using network science and quantitative modelling of exchange processes to investigate the genetic behaviours that these encounter networks could potentially support. 

\section*{Results}



\subsection*{The emergence of effective genomes on \emph{Arabidopsis} encounter networks as a network science problem}

\begin{figure*}
  \includegraphics[width=\textwidth]{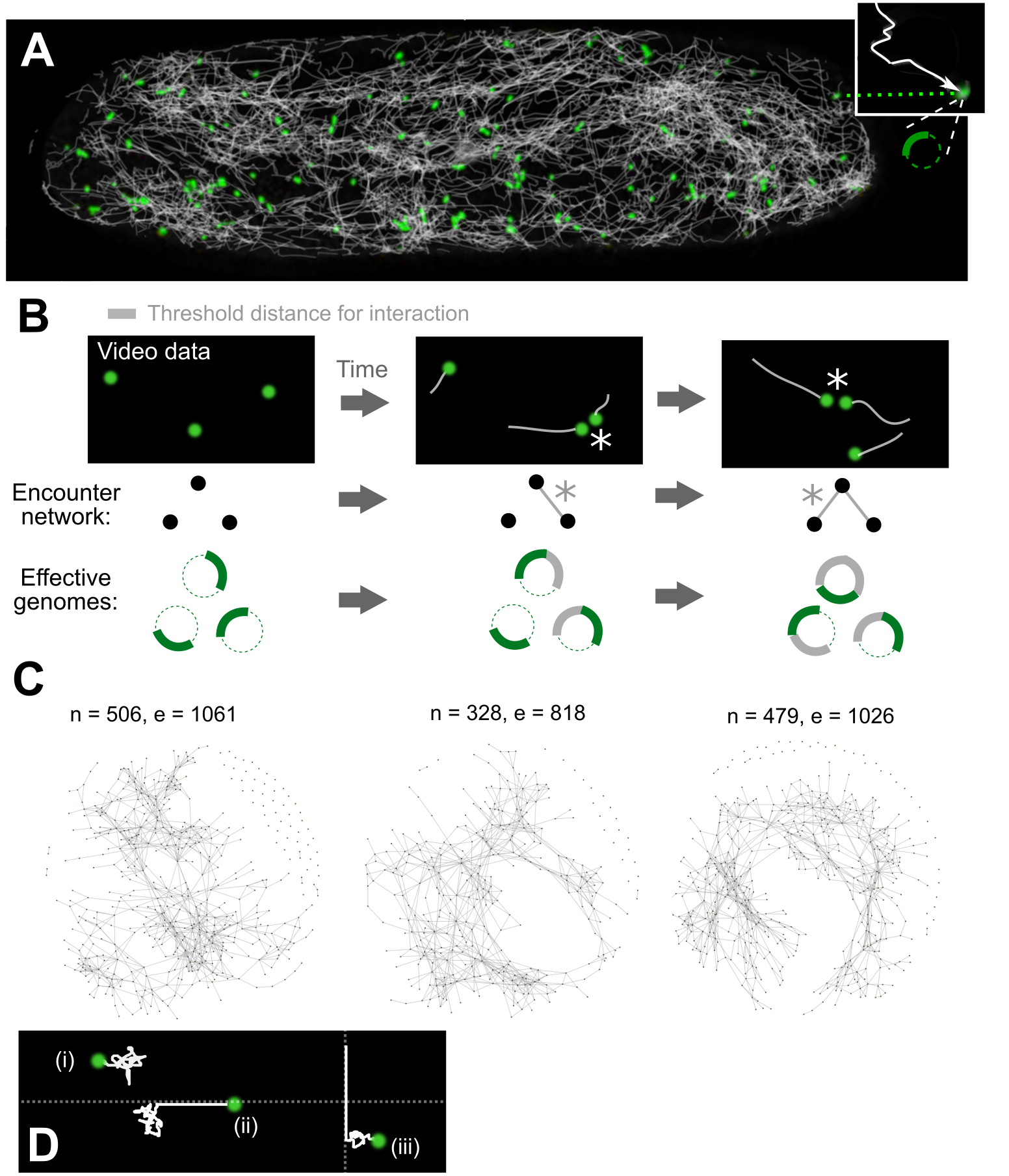}
  \caption{\textbf{Characterising mitochondrial encounter networks.} (A) Confocal microscopy with mtGFP \emph{Arabidopsis} \cite{logan2000mitochondria} creates videos of the motion of mitochondria (green) in hypocotyl cells. TrackMate \cite{tinevez2017trackmate} in Fiji \cite{schindelin2012fiji} is used to characterise trajectories (white; individual shown in inset). Individual mitochondria, as illustrated in inset, may only carry a reduced mtDNA molecule encoding a subset (thick line) of the full genome (dashed line).  (B) Trajectory sets are interpreted as encounter networks by representing each mitochondrion as a node, and connecting two nodes with an edge if they are ever colocalised within a given threshold distance ($*$). Encounters between mitochondria, if they lead to fusion and exchange of mtDNA, can expand the `effective genome' seen by a mitochondrion, as illustrated: green regions are the mtDNA molecule currently within a mitochondrion, grey regions are those that to which the mitochondrion has been recently exposed. (C) Example encounter networks constructed over a period of 231 seconds $n$ nodes, $e$ edges, $cc$ connected components). (D) Simple physical model used to simulate mitochondrial motion. Model mitochondria may (i) move purely diffusively with constant $D$; (ii) attach to the cytoskeleton with probability $k_{on}$ and then move ballistically; (iii) detach from the cytoskeleton with probability $k_{off}$ and continue to diffuse.}
  \label{fig1a}
\end{figure*}

We first sought to understand the process by which effective genomes could potentially emerge from dynamic interchange of subgenomic molecules in plant cells, using encounter networks characterised from hypocotyl cells in 7-day \emph{Arabidopsis} seedlings (see Methods). In previous work, we established an experimental and computation pipeline to characterise the `social' encounter networks of mitochondria \cite{chustecki2021network}. Here, nodes represent mitochondria, and an edge between two nodes means that those two mitochondria have colocalised within a physical threshold distance at at least one timepoint during the experiment (Fig. \ref{fig1a}). Example networks from mitochondrial dynamics in \emph{Arabidopsis} hypocotyl are shown in Fig. \ref{fig1a}. We will use these, and other experimentally-characterized encounter networks, in the subsequent analysis. Results from independent single cells were generally very similar (Fig. \ref{figsi-differentcells}).

We proceed by phrasing the core biological question -- the collection of genetic information by mitochondria -- as a network science problem. The biological problem is: how can individual mitochondria become exposed to the full mtDNA genome, given that each may only carry a reduced molecule? We will refer to an `effective genome' as the set of genes that a mitochondria has been exposed to over time.

\begin{figure}
  \begin{subfigure}{0.5\textwidth}
\fbox{\begin{minipage}{0.95\textwidth}Our network phrasing is as follows. Allow each node to have two labels $G$ (genome) and $H$ (history), both binary vectors of length $L$. $G$ describes the set of genetic elements within a node. $H$ describes the set of genetic elements that have been within a node at some point in the past. When a node's $H$-label contains $L$ elements of value 1, that mitochondrion has been exposed to every genetic element in the full set. 

Define an \emph{exchange event} between two nodes $a$ and $b$, connected by an edge, as follows. The $H$-label of $a$ acquires a 1 value at every element where the $G$-label of $b$ is 1. The $H$-label of $b$ acquires a 1 value at every element where the $G$-label of $a$ is 1. Then the $G$-labels of $a$ and $b$ are exchanged. Such an event corresponds to two mitochondria exchanging genetic information, with each being exposed to the genetic information currently in the other.

A given instance of the problem is defined by initial conditions ($G$-labels for each node) and an adjacency matrix. We are interested in how the $H$-labels of nodes (the sets of genetic elements that mitochondria have been exposed to) change as the number of exchange events increases. Following the nomenclature in the main text, the bingo score of a node is the proportion of 1s in its $H$-label $\sum_{i=1}^L H_i / L$, and a bingo is scored when this score is 1.
  \end{minipage}} \end{subfigure}
  \begin{subfigure}{0.5\textwidth}\includegraphics[width=\textwidth]{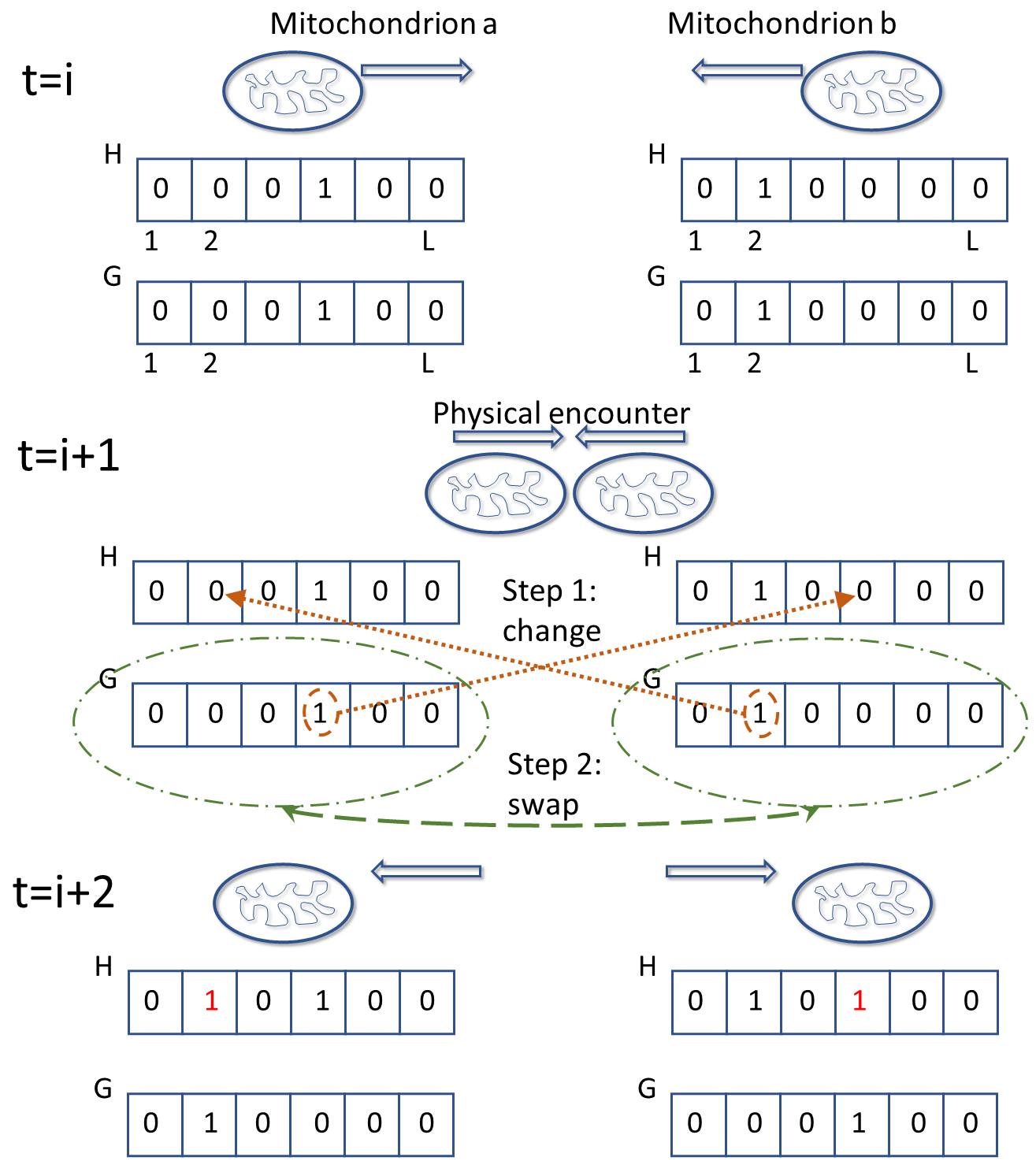}
    \end{subfigure}
\caption{\textbf{Effective genome emergence as a network science problem.} Outline of the algorithm modelling genetic exchange on encounter networks. Visual illustration gives an example of the corresponding cell biological process: at time $i$, two mitochondria with different genomes $G$ move towards each other. At time $i+1$, they undergo a physical encounter and exchange genomes. Their $H$ libraries (set of genes that have been seen at some time point) are updated with the new content their partner has provided. Their $G$ genomes are then exchanged. }
  \label{fig1}
\end{figure}

We present the specific phrasing of this problem in Fig. \ref{fig1}. Qualitatively, we ask how many genes an individual mitochondrion is exposed to over time, as a function of the proportion of encounters between mitochondria that lead to genetic exchange. We start with some initial state where each mitochondrion contains some or no genetic information, and investigate how information spreads through the population of mitochondria as exchanges between mitochondria occur. This problem shares structural similarities with a wide range of problems in epidemiology \cite{karp2000randomized,moore2000epidemics,kempe2004spatial,akdere2006comparison,chakrabarti2008epidemic}, probability theory (including variants of the coupon collector problem \cite{flajolet1992birthday,newman1960double}), communication networks and algorithms (including the requirement for every node in the network to acquire required information about the existence of their neighbors \cite{vasudevan2009neighbor,ye2012efficient}), but has some key differences (see Discussion).


For brevity, we refer to this as the \emph{bingo problem}, by analogy with the collection of a set of elements which is built up over time. A node's \emph{bingo score} is the proportion of genetic elements that it has been exposed to over time. A \emph{bingo} occurs when a node has a bingo score of one. This corresponds to a mitochondrion having been exposed to the full set of elements in the genome. An informative summary of a given cell's performance is the proportion $p$ of nodes that have scored bingos (the proportion of mitochondria that have been exposed to a full effective genome).

While a physical encounter does not necessarily imply fusion and exchange of genetic content, it is a requisite for this exchange. We therefore consider how effective genomes emerge as a changing proportion of encounters are interpreted as leading to exchanges. On one hand, if no encounters lead to exchanges, effective full genomes will never emerge. On the other, if every encounter leads to an exchange, effective full genomes may emerge readily. To characterise this behaviour, we simulated effective genome emergence via the `bingo' game in Fig. \ref{fig1}. We recorded the proportion $p$ of nodes that have scored a bingo (the proportion of mitochondria that have experienced a full effective genome) as a function of the proportion $q$ of encounters that correspond to an exchange. We increase $q$ following the temporal ordering of encounters in the network.

\begin{figure*}
  \includegraphics[width=0.5\textwidth]{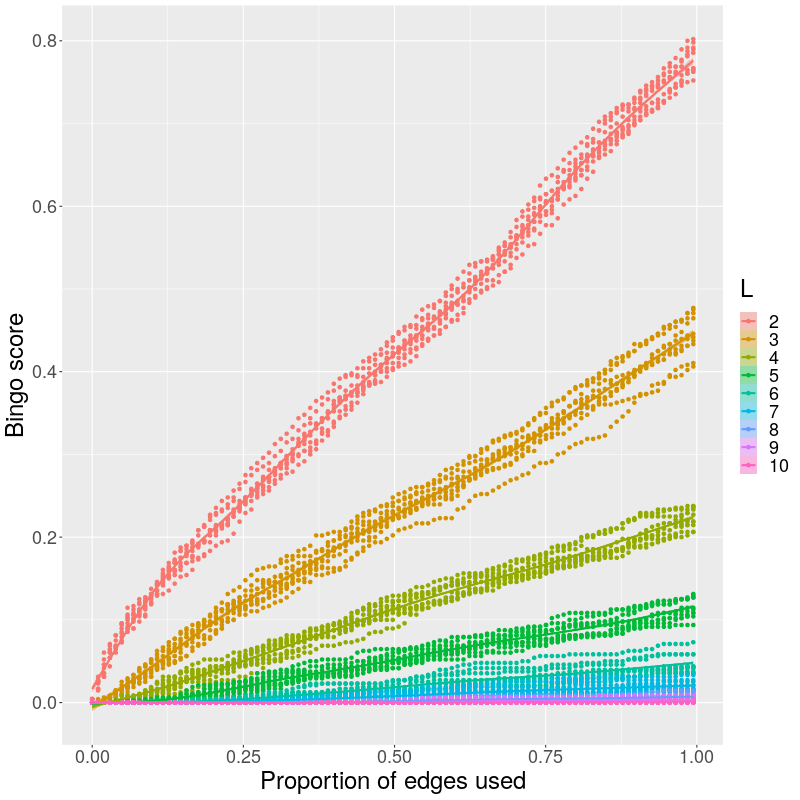}   \includegraphics[width=0.5\textwidth]{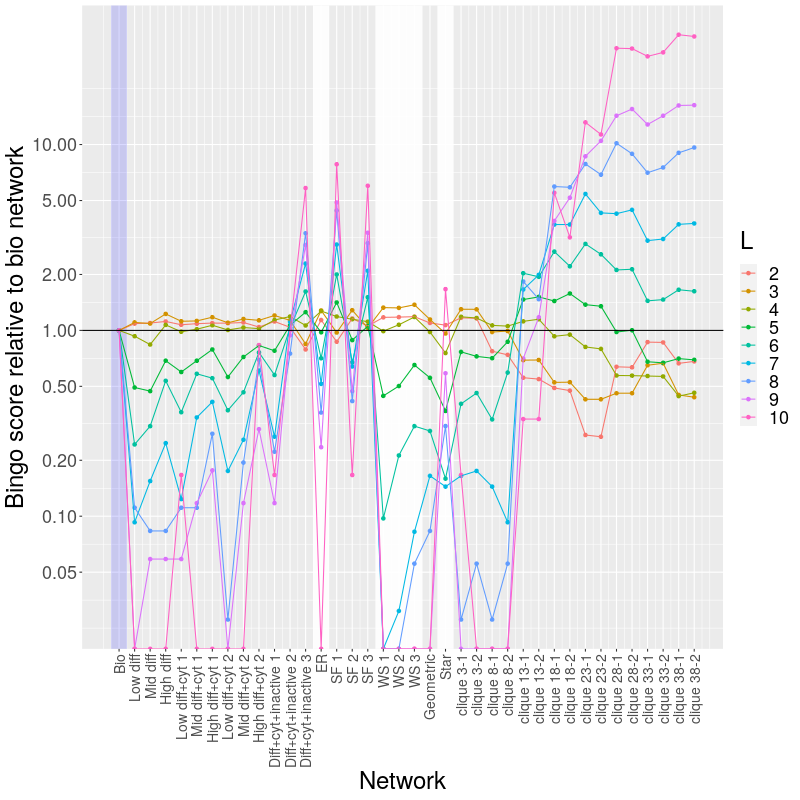}
\caption{\textbf{Potential for genome emergence on \emph{Arabidopsis} encounter networks and on different theoretical network structures.} (A) The `bingo score' (proportion $p$ of mitochondria that have experienced a full effective genome), as a function of the proportion $q$ of encounter network edges (physical encounters) that allow genetic exchanges. As $q$ increases, genetic information spreads through the mitochondrial population, and more individuals `collect' the full set of genetic information, increasing $p$. This increase depends strongly on $L$, the number of different genetic elements that constitute the full effective genome: higher $L$ means more elements must be `collected', which requires correspondingly more information exchange. Ten simulations were performed for each $L$ value, using an experimentally-characterised \emph{Arabidopsis} encounter network (see text). (B) Final bingo score $p^*$, the proportion of mitochondria that have experienced a full effective genome if all encounters allow genetic exchange, is computed for all graph types. This plot shows $p^*/p^*_0$, this quantity normalised by the value for the biological network structure. For low $L$, some theoretical networks outperform biology but cliquey networks perform poorly. For high $L$, the situation is reversed. Traces connecting different network results are drawn to reflect the profile of results for a given $L$ and do not reflect any relationship between different networks.  Networks immediately to the right of `Bio' are encounter networks from physical simulation; others from synthetic construction. Labels: \emph{diff}, diffusion; \emph{cyt}, cytoskeletal motion; \emph{inactive}, stochastic inactivation of mitochondria (modelling entering and leaving the domain); \emph{ER}, Erd\H{o}s-R\'{e}nyi; \emph{SF}, scale-free; \emph{WS}, Watts-Strogatz; \emph{clique x-y}, graph with cliques of size $x$, disconnected if $y = 1$ or connected by a single edge if $y=2$. Different network classes appear on alternating grey and white backgrounds.}

    \label{fig2a}
\end{figure*}

Intuitively, the dynamics of genome emergence depend strongly on $L$, the number of different genetic elements that are required to make up a full effective genome (Fig. \ref{fig2a}A). For low $L=2$, effective genomes rapidly emerge with low numbers of interactions, and in the $q=1$ case where all edges lead to exchange, a majority of mitochondria are able to collect a full effective genome. For higher $L$, collection becomes increasingly challenging, with only around $10\%$ of mitochondria collecting a full effective genome with $q=1$ and $L=5$, and fewer for higher $L$.

\subsection*{\emph{Arabidopsis} encounter networks support efficient emergence compared to theoretical encounter networks}

Having characterised the potential for effective genome emergence on \emph{Arabidopsis} encounter networks, we next asked how these biological networks compared to theoretical alternatives in their capacity to support such emergence. To this end, we investigated the bingo problem on a set of synthetic encounter networks.

For each experimentally-characterised network, we built a range of synthetic networks constrained to have the same numbers of nodes and edges (Fig. \ref{figsi-graphs}). Our theoretical networks began with Erd\H{o}s-R\'{e}nyi (ER) random topologies (\cite{erdHos1960evolution}; edges placed between pairs of nodes randomly chosen with uniform probability), scale-free (SF) topologies (\cite{barabasi1999emergence}; edges placed between pairs of nodes randomly chosen with probability proportional to their degree), and Watts-Strogatz (WS) networks (\cite{watts1998collective,moore2000epidemics}; a `ring-like' network with subsequent rewiring to reduce networks distances).

We further explored several other network types: geometric random graphs \cite{penroseRGGs}, star graphs, and `cliquey' graphs. The final class followed our hypothesis that `cliquiness' in networks would more directly lead to efficient genome emergence, as follows. Cliquey networks consist of cliques (sets of nodes that are all mutually connected) with few or no connections outside each clique. Nodes within cliques can then rapidly assimilate all available genes without risk of `losing' them to a broader set of partners. We constructed two classes of cliquey network: (i) disconnected cliques of size $n$ and (ii) cliques of size $n$ connected by a single link. In each of these synthetic cases, we specified a number of nodes to match a biologically observed network and padded the network with random edges if necessary to match that network's edge count.

We found that the bingo performance of different networks depends strongly on $L$, with some networks performing relatively well at $L \leq 3$ (ER, WS, geometric, small cliques) and poorly at $L \geq 4$, and some with the opposite pattern (larger cliques) (Fig. \ref{fig2a}B, Fig. \ref{figsi-differentcells}).

This picture immediately suggests a tension in clique size. Smaller cliques will share information more rapidly. But if a clique is too small, it may not possess all the genes required to accumulate the full set. We found that for $L = 2$, bingo performance was a simple function of clique size, with smaller cliques (down to $n_c = 3$) performing best, and larger cliques (up to $n_c = 38$) performing worst. However, as $L$ increased, this picture became more nuanced. For $L = 3$, the performance of $n_c = 3$ networks was substantially challenged, due to the probability of a clique not possessing a copy of each genetic element. For $L=3$, larger cliques ($n_c = 8$) performed better, with even larger clique sizes ($n_c$ between 10 and 25) performing best for higher values of $L = 4$ to $L=6$. Larger cliques $n_c > 30$ performed poorly in most cases, only becoming broadly competitive at high $L$ values.

However, the more striking result was that biological networks and SF networks were the most robust performers. While never being the best performer for a given $L$, these networks performed much more consistently across a range of different $L$ values (Fig. \ref{fig2a}B).



\subsection*{Heterogeneous diffusive and ballistic motion supports efficient effective genome emergence}

We next asked which properties of biological mitochondrial motion were responsible for the formation of encounter networks with strong bingo performance. To this end, we considered a simple physical simulation following \cite{chustecki2021network} (Fig. \ref{fig1a}D; Methods). Within the simulation, mitochondria move diffusively, with some probability of attaching to a cytoskeletal strand, whereupon they move ballistically until they detach with some probability. The attachment-detachment probabilities, diffusion constant, and speed when attached to a strand are parameters of the simulation.

Exploring a range of parameters in this model (see Methods), we found that no instance of the diffusive-ballistic model produced encounter networks that could outperform biological networks at bingo. While simulated performance was marginally higher for $L \leq 3$, performance at higher $L$ was substantially lower, only approaching the biological case for unphysically high values of the diffusion constant and ballistic speed (Fig. \ref{fig2a}B). The degree distributions of networks constructed through simulation typically had more limited spread, with fewer nodes of high degree (Fig. \ref{figsi-degrees}).

We and others previously observed pronounced inter-mitochondrial heterogeneity in dynamics. Some mitochondria persist in a given cellular region for a long time period, whereas others enter and leave the region, leading to heterogeneity in the time windows for which a given mitochondrion is present. Those individuals present for longer have more opportunity to encounter partners and become highly connected. To model this, we introduced another process in our simulation model, allowing mitochondria to enter and exit the region of observation randomly with given rates (see Methods). As before, we used simulations to produce encounter networks matching the node and edge count of the biological original. We found that these simulated networks, with high diffusion and cytoskeletal motion, more resembled the biological bingo performance (Fig. \ref{fig2a}B). Hence, a combination of diffusive and ballistic motion with broader variability in individual behaviour builds a foundation for efficient genome emergence.

To further explore this observation, we next artificially truncated the length of tracked trajectories in the biological data. Unsurprisingly, this led to smaller encounter networks, but also amplified the performance boost of scale-free and beneficially cliquey networks (Fig. \ref{figsi-differentexpts}). This observation supports the picture where a subset of individuals, remaining in the system for a comparatively long time period, accumulate more encounters and thus help facilitate the beneficial exchange of contents.

\subsection*{Network properties linked to efficient effective genome emergence}
Given these observations, we next asked whether simple summary statistics of network structure correlated with bingo performance, and hence whether particular structural features might conceivably be selected in cellular control of mitochondrial encounter networks. It may be anticipated that a network's performance at bingo would be related to how rapidly information can be spread through the network. This rapidity is captured by statistics like the global network efficiency $\nu = (n(n-1))^{-1} \sum_{i\not= j \in G} d(i,j)^{-1}$, the sum of the reciprocals of shortest path lengths $d(i,j)$ between all pairs of nodes $i$ and $j$, normalised by the number of pairs $n(n-1)$. Structural statistics like modularity (which we measure here using the walktrap algorithm \cite{pons2006computing}) and the size and structure of connected components may also be anticipated to play a role (the mean degree, by construction, is equal across all networks compared in an experiment).

However, when exploring bingo behaviour on our synthetic networks, we found that networks with high efficiency, and high values of other intuitively desirable statistics, often do not perform well at bingo (Fig. \ref{figsi-correlations}). It is in every node's interest to be the only node connected to as many other sources of information as possible; efficient networks typically connect `everything to everything'. Other summary statistics also failed to show a tight correlation to bingo performance. While some correlated strongly for a given $L$ (for example, increasing number of connected components decreases performance for $L = 2$), these relationships were typically reversed for different $L$ (increasing number of connected components increases performance for $L = 5$). One suggestive observation is that those networks that perform most consistently -- SF and biological networks -- have a high degree `range', here defined as the number of values $k$ for which at least one node in the network has degree $k$ (Fig. \ref{figsi-degrees}). This quantity is at least somewhat related to the `scale-free' nature of a network -- degrees spanning a wide range of values -- perhaps suggesting the capacity to accumulate information over a diverse ranges of `scales' of $L$.

Given this observation, we next considered a more concrete theoretical framework to understand the problem of effective genome emergence -- specifically, the coupon collector's problem or CCP \cite{ferrante2014coupon}. The informal phrasing of the problem is: if each cereal box contains a random coupon, and there are $n$ different types of coupon, how many cereal boxes do I need to buy to collect all $n$ types? The CCP generally describes the process of sampling coupons (which are individual members of a set of coupons $\mathcal{L}$) from a certain number $n$ of `urns' (entities containing coupons) $n$. In our system, coupons correspond to individual genome regions (members of the full genome), and urns correspond to mitochondria containing these genes (to further draw the analogy between the CCP and the bingo game for effective genome, we refer to the visualised glossary in Supp. Fig. \ref{figsi-bingoAnalogy}.) We consider the CCP faced by an individual mitochondrion -- a node $s$ in our encounter network. This node begins with its initial gene, and through encounters can draw from each its neighbours (of which there are $deg(s)$). So its total number of draws is $n(s) = deg(s) + 1$, and the number of distinct coupon types to collect is $| \mathcal{L}| = L$.





Study of the CCP has answered many questions about this system -- some examples linked to this system appear in Refs. \cite{flajolet1992birthday, adler2003coupon, schilling2021results}. The most central for us is, given $n$ draws, what is the probability of collecting all $L$ coupons? A classical result, outlined in Methods, is that

\begin{equation}
  P(s\,\, \text{scores bingo}) = \sum^{L}_{j=0} (-1)^{j}{L \choose j}\left(\frac{L-j}{L}\right)^{n(s)},\label{bingoAnalysis}
\end{equation}

The expected number of neighbours required to score a bingo is also easily derived (see Methods) to be

\begin{equation}
  E(n(s) | \text{$s$ scores bingo}) = LH_{L},\label{bingoAnalysisExpectedDer}
\end{equation}
where $H_L$ is the $L$th harmonic number.

\begin{figure*}
  \includegraphics[width=\textwidth]{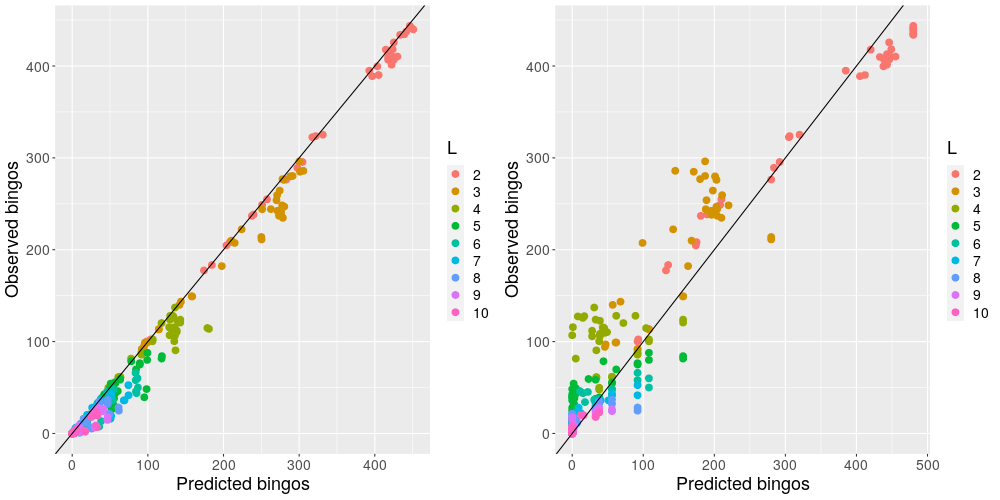}
\caption{\textbf{Analytic results predict bingo performance on biological and synthetic encounter networks.} Predicted (horizontal axis) and observed (vertical axis) number of nodes scoring bingo; each point corresponds to a different network. (left) expected results from Eqn. \ref{bingoAnalysis}; (right) estimated performance using the threshold value derived from the expected number of encounters needed to fill a bingo. There are 20 repetitions of each bingo game with different $L$, each value of $L$ is represented with a different colour.}
  \label{analysisPlotScatThres}
\end{figure*}

Given this quantity, we are able to characterize and `predict' the behaviour of a graph structure in bingo, including mitochondrial encounter networks, based on a simple scalar property of the network. Fig. \ref{analysisPlotScatThres} confirms that the bingo game corresponds to the CPP variation described in Eq. \ref{bingoAnalysis}. We see that the equation predicts the game's outcome for the majority of network topologies and across different values of $L$; the approximate prediction using the expected value given by Eq. \ref{bingoAnalysisExpectedDer} also reasonably predicts the bingo outcome, while requiring only a summary statistic of the whole network.

These insights support the intuitive observation that nodes with degree less than $L$ can never score a bingo, and thus have a purely negative effect on the bingo performance of a network when measured by the proportion of bingo scores. Such nodes, including `singletons' with degree zero, do occur in our biological encounter networks, because of the limited time window of our observation (see Discussion). To check how much our general results depend on the presence of these low-degree nodes, we artificially removed degree-zero nodes from our biological encounter networks, and re-analysed these `pruned' networks as above, constructing new synthetic and simulated networks to match the new node and edge counts. We confirmed that networks with `pruned' and original statistics showed very comparable behaviours, showing that the typically small proportion of singletons does not dramatically influence overall network performance (Fig. \ref{figsi-differentexpts}.)


\subsection*{Presence of `master circles'}

The previous sections considered a population of mitochondria where each mitochondrion begins with one genetic element. An alternative picture in plant biology is that of `master circles' and `small circles'. Here, some mitochondria possess full copies of the mtDNA genome, and many mitochondria possess reduced copies. Those with full copies have been pictured as `genetic vaults' or `repositories' of genetic information \cite{logan2010mitochondrial,johnston2019tension}.

We next investigated how the presence of master circles influences the emergence of an effective genome across mitochondria. Clearly, the dynamics of the bingo game will differ, because any mitochondrion with a master circle immediately attains a score of one. We explored the behaviour of the system when $1\%$ or $2\%$ of the mitochondrial population was initialised with a master circle, denoting by $m$ the proportion of mtDNA molecules that are master circles. Intuitively, we saw higher bingo scores over time in the cases where more master circles were present (Fig. \ref{figsi-bingom}). We also saw a decrease in the scale of differences between networks, with biological and theoretical networks performing more similarly. Interestingly, the presence of master circles induces non-monotonic behaviour of performance with $L$. In contrast with the $m = 0$ case, where high $L$ values corresponded to maximal cliquey performance and minimal simulation performance, here cliquey performance is maximised and theoretical performance minimised around $L=5$ or $L=6$, with the trend reversing at both higher and lower $L$. The dynamics of bingo is more comparable across different networks for $m > 0$ (Fig. \ref{figsi-dynamicsm}), with biological networks performing generally well across all $L$. Hence, while $m > 0$ makes the bingo problem generally easier for all networks, and biological networks perform correspondingly well.

\section*{Discussion}
The previous research presenting these encounter networks \cite{chustecki2021network} hypothesised that the collective dynamics of plant mitochondria allow the cell to balance two priorities. The first is an even physical distribution of mitochondria, ensuring a uniform energy supply, potential for colocalisation with other organelles throughout the cell, and avoiding heterogeneity in concentration of metabolites and signalling molecules. The second `social' priority is colocalisation of mitochondria to facilitate exchange of genetic information and biomolecules. Here we build on this second priority to show that the topology of encounter networks is capable of facilitating the efficient emergence of an effective genome.

We have shown that the encounter networks observed in plant cells support the emergence of a full `effective' mtDNA genome in an efficient and distributed way. We showed that biological encounter networks support this process more than a range of synthetic and simulated networks, demonstrating that this process is an analogy of the coupon collector's problem (CCP) and characterising it mathematically.

The above explains some interesting observations from the empirical part of the paper:\\
\emph{Scale-free networks demonstrated good bingo performance across $L$.} This happens because of the well-known heavy tails in their degree distribution. This enables them to perform adequately well for large $L$ and at the same time they are efficient for small $L$, as well (showing high degree of robustness \cite{liu2017comparative}).\\
\emph{Biological networks perform relatively well at bingo.} The degree distribution of biological encounter networks is quite similar to the degree distribution of the scale-free networks (for the same given number of nodes and edges). This enables them to have a tail almost the same as the scale-free cases. This is not new, since mitochondria are seen to form \emph{social}-like networking structures \cite{chustecki2021network} and the results here seem to further validate those findings.\\
\emph{Cliquey networks have an `unstable' performance for different $L$.} Because for large values of $L$ and big clique size they act as an approximation of a complete graph, which in turn has the maximum degree distribution and hence it is the optimum topology to play bingo (if given of course a much higher number of edges) \cite{aldous1989introduction}. On the other hand, they fail to perform well for small $L$ because they have `sacrificed' big chunks of the network edges to form the cliques. The exact opposite behaviour is observed for small cliques, where the performance is maximized for low values of $L$, whereas for $L>5$ they do poorly.\\
\emph{Other networks (diffusion, random, and so on) do reasonably for small $L$, but less well for larger ones.} Because their degree distribution lacks a tail needed to expect successful bingos. For example, for $L=8$, 21 neighbors are expected in order to have a complete bingo with high probability.


We have not considered mtDNA replication, degradation, recombination, or other genetic dynamics \cite{johnston2019tension} in this model. Plant mitochondrial DNA readily recombines (unlike animal mitochondrial DNA), allowing mixing and restructuring of the information shared between mtDNA molecules \cite{woloszynska2009heteroplasmy}. Here we only consider the question of mitochondrial access to genetic information, not the population dynamics and/or restructuring of the molecules containing this information. This is a rich topic in itself, addressed by some classical \cite{atlan1993model, albert1996dynamics} and some recent \cite{edwards2021avoiding} theory, and the influence of these physical dynamics of mitochondria on the genetic dynamics of mtDNA is an ongoing topic of research \cite{mogensen1996hows, mouli2009frequency, poovathingal2009stochastic, johnston2019tension, aryaman2019mitochondrial, tam2013mathematical, tam2015context, hoitzing2017stochastic}. We underline that the details of rates and magnitudes of our proposed mechanism remain hypothetical: although elegant experiments have demonstrated contents exchange and mixing throughout the chondriome \cite{arimura2004frequent, arimura2018fission}, the physical and temporal scales of inter-organelle mtDNA exchanges remain, to our knowledge, uncharacterised. Experimental characterisation of these processes will allow parameterisation of our model, which for now demonstrates the range of possible behaviours and general principles without specifying given parameter values.

Like any approach based on imaging, our characterisation of biological encounter networks is subject to some noise. The requirements to image the cell with a fine time resolution (so that mitochondria can be accurately tracked) and with limited laser power (to avoid damaging the cell) limit the resolution of individual frames, and the motion of mitochondria, while largely confined to a 2D plane, can sometimes lead to individuals being lost during the tracking process. This can affect the structure of the subsequent encounter networks. However, the most common issue -- a mitochondrion being transiently `lost' and hence, for example, being represented as two mitochondria (before and after the `loss') early and late -- will generally have the effect of reducing the degree of nodes. This is because the set of encounters of such a mitochondrion will be split between the two individuals. We thus expect the `true' encounter network to involve more higher-degree nodes, thus supporting the distinction from the synthetic cases with limited degree distributions. On a similar note, our protocol involves imaging over a finite time window. Over time, encounter networks will gain more edges, and it is conceivable that over a long time the networks will come to resemble a complete graph, with every mitochondrion having encountered every other. However, there is another timescale in the system: the timescale on which genetic information is `forgotten', as protein products expressed from a historically-encountered genome molecule degrade. The system is thus expected to avoid steady state behaviour, and our approach informs about the dynamics that shape the system in a sampled window of this out-of-equilibrium behaviour. Further, plant cells are dynamic systems capable of responding to internal and external stimuli via sensing and feedback control. As such, the topology of a cell's encounter network is not fixed over the lifetime of the cell. Cells may adapt mitochondrial dynamics to favour, for example, `cliquier' or sparser encounter networks as circumstances demand. The capacity of the cell to control mitochondrial dynamics to optimise mitochondrial exchange, and other priorities, is an exciting target for future work.

Our bingo problem resembles many questions from the field of dynamic networks, found in other fields \cite{moore2000epidemics,vasudevan2009neighbor,flajolet1992birthday,cao2018panacea}. One particular feature of our plant system is that information cannot be duplicated (mtDNA molecules are assumed not to replicate over the timescale of these dynamics). Once a mitochondrion has been exposed to an element, it remembers that exposure, but can only pass on the information from that element if it possesses an mtDNA molecule including it -- whereupon it loses that molecule. 

In conclusion, we have shown that the dynamic encounter networks of mitochondria in \emph{Arabidopsis} cells have the capacity to support efficient mtDNA complementation, allowing individual mitochondria to `collect' an effective genome despite only ever carrying a reduced subset. Under several circumstances, this genome emergence seems more efficient in biological networks than in many theoretical cases which may be expected to perform well. This suggests an intriguing hypothesis -- that the cellular control of these encounter networks may have evolved to facilitate efficiency genome emergence. If this is indeed the case, plant mitochondrial dynamics represents a `social network' structure under evolutionary control to fulfil an important cellular function. 

\section*{Methods}
\textbf{Plant growth.} (experimental protocols follow those in Ref. \cite{chustecki2021network}). Seeds of \emph{Arabidopsis thaliana} with mitochondrial-targeted GFP (kindly provided by Prof. David Logan \cite{logan2000mitochondria}) were surface sterilized in 50\% (v/v) household bleach solution for 4 minutes with continual inversion, rinsed three times with sterile water, and plated onto $\frac{1}{2}$ Murashige and Skoog (MS) agar. Plated seeds were stratified in the dark for 2 days at 4${}^{\circ}$C. Seedlings were grown in 16hr light/8hr dark at 21${}^{\circ}$C for 4-5 days before use.

\textbf{Imaging.} Prior to mounting, cell walls were stained with 10$\mu$M propidium iodide (PI) solution for 3 minutes. Following a protocol modified from \cite{whelan2015plant}, full seedlings were mounted in water on microscope slides, with cover slip. Imaging of dynamic systems in living cells is a balance between spatial/temporal resolution and maintaining physiological conditions. To avoid undesirable perturbations to the system including physical and light stress and hypoxia, all imaging was done maintaining low laser intensities and within at most 10 minutes of mounting to minimise the effects of physical stress and hypoxia (Prof Markus Schwarzl\"{a}nder, personal communication).

A Zeiss 710 laser scanning confocal microscope was used to capture time lapse images. To test robustness of the imaging protocol, a Zeiss 900 with AiryScan 2 detector was also used for several identically prepared samples, with no differences between summary statistics collected from these samples and those from the 710 beyond natural variability. For cellular characterisation we used excitation wavelength 543nm, detection range 578-718nm for both chlorophyll autofluorescence (peak emission 679.5nm) and for PI (peak emission 648nm). For mitochondrial capture we used excitation wavelength 488nm, detection range 494-578nm for GFP (peak emission 535.5nm). Videos were 231 seconds long, with a frame interval of 1.94 seconds, and a resolution (after scaling for standardisation) of 0.2 $\mu$m per pixel.

\textbf{Video analysis.} Individual cells were cropped from the acquired video data using the cell wall PI signal using Fiji (ImageJ) \cite{schindelin2012fiji}. The size of each video was scaled to the universal length scale 5.0 pixels/$\mu$m. We then extracted individual mitochondrial trajectories from the acquired video data using TrackMate \cite{tinevez2017trackmate}. Typical settings used were application of the LoG Detector filter with a blob diameter of 1$\mu$m and threshold of 2-7, filters were set on spot quality if deemed necessary. The Simple LAP Tracker was run with a linking max distance of 4$\mu$m, gap-closing distance of 5$\mu$m and gap-closing max frame gap of 2 frames. In each case we visually confirmed that individual mitochondria were appropriately highlighted and that tracks were well captured, editing occasional tracks where necessary. XML output from TrackMate was converted to adjacency matrices using custom code (see below).


\textbf{Null model networks.} We constructed several theoretical models for network structure, each with $n$ nodes and $e$ edges. First, Erd\H{o}s-R\'{e}nyi (ER) random networks \cite{erdHos1960evolution} were constructed by randomly choosing two non-identical nodes $a$ and $b$, each with probability $1/n$, and creating an edge between them, repeating until $e$ edges were created.

Second, scale-free (SF) networks \cite{barabasi1999emergence} were constructed by randomly choosing nodes with probability $1/deg(a_i)+1 / \sum_j 1/deg(a_j) + 1$. This procedure was repeated $e$ times, with degree updated each time, for the basic network (i). Variations of scale-free networks were created in two ways. For (ii), beginning with a linear network where an edge connects each $a_i$ and $a_{i+1}$, then proceeding as in (i), thus enforcing connectivity. For (iii), a preferential attachment process was performed for each of the $n$ nodes, where a node is connected to a partner $a$ with probability $1/deg(a_i)+1 / \sum_j 1/deg(a_j) + 1$, where the sum $j$ is over nodes added so far to the network. Extra edges are then added as in (i).

Third, Watts-Strogatz (WS) networks \cite{watts1998collective} were constructed as follows. Compute the mean degree $k = n/e$. Label each of $n$ nodes with successive integers. For each node $i$, draw $k_i =  \lceil k \rceil$ or $\lfloor k \rfloor$ randomly with relative probabilities $\lceil k \rceil -k$ and $k - \lfloor k \rfloor$. If $k_i$ is even, connect $i$ to the $k_i / 2$ nodes immediately before it and the $k_i / 2$ nodes immediately after it in sequence. If $k_i$ is odd, connect to $(k_i+1)/2$ `before' nodes and $(k_i-1)/2$ `after' nodes with probability $\frac{1}{2}$, or vice versa with probability $\frac{1}{2}$. For all edges linking $i$ to a node with label $>i$, change the target node with probability $\beta$ to a different node $\not= i$.

Fourth, `cliquey' networks were constructed. Given a clique size $c$ and constraints on $n$ and $e$, the number of cliques allowed was computed as $n_c = \min (\lfloor n/c \rfloor, \lfloor e / (c(c+1)/2) \rfloor )$. The $n$ nodes were partitioned into $n_c$ cliques with edges between each pair of nodes within each clique. These cliques were then either (i) left disconnected; (ii) connected with a single edge linking two cliques; (iii) left disconnected but padded with randomly placed edges to reach $e$ total; (iv) connected with a single edge linking two cliques then padded.

Fifth, geometric random graphs (GRGs) were constructed by placing $n$ points -- each representing a node -- in the unit square, and progressively adding edges between the two disconnected nodes with the shortest distance between their corresponding points, until $e$ edges existed. Finally, the star graph with $n$ nodes was constructed by connecting $n-1$ nodes to a central node, then adding random edges until $e$ edges existed.

\textbf{Model networks based on physical simulation.} Synthetic encounter networks were constructed based on physical simulation of model mitochondrial dynamics using custom code in C (see below). As we are free to set length and time units in our simulation, we use $1 \mu m$ as the unit of length and set one discrete simulation timestep equivalent to $1s$. $n$ agents were simulated in a model cell, a 2D rectangular domain with reflecting boundary conditions at $x = 0$, $x = 100 \mu m$, $y = 0$, $y = 30 \mu m$, to model the geometry observed in our experimental observations of hypocotyl cells \cite{chustecki2021network}. Cytoskeleton strands are modelled as crossing the cell at constant $x$ (horizontal) and at constant $y$ (vertical). Each agent could, at any time point, be detached or attached to the cytoskeleton. If detached, each timestep, agents were moved according to a normal kernel with standard deviation $2D \mu m^2 s^{-1}$, so that $D \mu m^2 s^{-1}$ is the diffusion constant. When first attached, an agent is assigned a velocity vector: while attached, that agent moves by that vector each timestep. The velocity vector is randomly chosen on attachment and may be in the $+x, -x, +y,$ or $-y$ direction, and has magnitude $V \mu m s^{-1}$. Each timestep, detached agents become attached with probability $k_{on}$, and attached agents become detached with probability $k_{off}$, corresponding to rates of $k_{on/off} s^{-1}$. When two agents were present within a distance $1.6\mu m$ of each other, an edge corresponding to the pair was added to the encounter network (if not already present). The physical simulation proceeded until $e$ edges were present.

Characteristic values observed experimentally are $D \simeq 0.1 \mu m^2 s^{-1}$ and $V \simeq 1 \mu m s^{-1}$ \cite{chustecki2021network}. In our simulations we explored one order of magnitude either side of these values, using $D = (0.02, 0.1, 1) \mu m^2 s^{-1}$ and $V = (0.1, 1, 10) \mu m s^{-1}$. We explored $(k_{on}, k_{off})$ pairs of $(0,0) s^{-1}$ (no cytoskeletal motion), $(0.1, 0.1) s^{-1}$, and $(0.5, 0.1) s^{-1}$.

Entry and exit of individual organelles into the system was modelled by switching individuals between `active' and `inactive' states. Active mitochondria behave as above and interact; inactive mitochondria remain static and do not contribute to any encounters, remaining effectively invisible (thus having exited the system). When this feature was used in simulations, activation and inactivation of individuals were stochastic events with rates $\rho_{on} = 0.01 s^{-1}$ and $\rho_{off} = 0.1 s^{-1}$ respectively, leading to a mean of 10\% active mitochondria at a given time.

\textbf{Coupon collector's problem.} Consider the different patterns of coupons that can be acquired through $n$ draws. There are $L^n$ possible patterns, which we assume all arise with equal probability. We require the probability of obtaining a pattern in which each of the $L$ coupons is present. To get this we use the inclusion-exclusion principle.

We first write down the probability of obtaining a pattern that is compatible with there being an `alphabet' of $L$ coupons. The probability of a single draw being compatible with an alphabet of $L$ coupons is $L/L = 1$, so we begin with a probability of $1^n = 1$. We need to deduct the probability of obtaining a pattern that is compatible with there being an alphabet of $L-1$ coupons, because every such pattern cannot feature all $L$ coupons. The probability of an individual draw not obtaining a given coupon $l$ is $(L-1)/L$, so considering each $l \in \mathcal{L}$ we obtain $L\times ((L-1)/L)^n$.

However, we have now over-counted patterns that are compatible with an even smaller alphabet size of $L-2$. So we need to add back the patterns that we have missed. The probability of an individual draw not obtaining either of a given pair of coupons $(l_1, l_2)$ is $(L-2)/L$, so considering each pair of coupons $(l_1, l_2) \in \mathcal{L}$ we have $\binom{L}{2} \times ((L-2)/L)^n$.

However, we have now over-counted patterns that are compatible with an even smaller alphabet size of $L-3$. We thus need to consider triplets of excluded coupons, and so on. The process continues iteratively, alternating between adding and subtracting terms (including and excluding) until we reach $L$ terms. From the above it should be clear that the final form is
\begin{eqnarray}
  P(\text{bingo}) & = & 1 - L \left( \frac{L-1}{L} \right) ^n + \binom{L}{2} \left( \frac{L-2}{L} \right)^n - ... \\
  & = & \sum_{j = 0}^L (-1)^j \binom{L}{j} \left( \frac{L-j}{L} \right)^n
\end{eqnarray}

For example, consider $n = L = 3$. Write the 24 three-character strings of length 3 for the set of patterns: AAA, AAB, .... At the first step we include them all. The next step counts all the strings that do not contain A, all those that do not contain B, and all those that do not contain C. Hence, we remove BBC, BCC, and so on -- but AAA, BBB, and CCC each get double-counted (once for each coupon they do not contain). The third step recounts all the strings that do not contain A or B, those that do not contain B or C, and those that do not contain A or C, which are exactly those three strings we previously double-counted. As $L=3$, this is our final step, and we have successfully retained only those strings in which all coupons feature.

The expected number of draws required for a bingo is easier to compute. If we have collected $c$ coupons, the probability of the next draw obtaining an unseen coupon is $(L-c)/L$. Assuming that draws are Bernoulli trials, a geometric distribution describes the behaviour of the system, giving a mean number of draws $L/(L-c)$ required for the next unseen coupon. The expected overall number is then $\sum_{c = 0}^L L/(L-c) = L ( 1/L + 1/(L-1) + ... + 1/1) = LH_L$.

It is important to note that the above equations assume an equal probability for each draw, as this is the case in the bingo game where $L$ values are initially scattered uniformly across the mitochondria/nodes. The influence of the coupon distribution for the same problem is beyond the scope of this work and is still an open topic of active research. The interested reader is referred to the work of Shilling \cite{schilling2021results}.

\textbf{Source code and data availability.} All source code is available at \url{github.com/StochasticBiology/mito-network-sharing}. We used TrackMate \cite{tinevez2017trackmate} in Fiji \cite{schindelin2012fiji} for video analysis, C with the igraph library \cite{igraph} for network generation and simulation, and R \cite{rsoftware} with libraries igraph \cite{igraph}, brainGraph \cite{braingraph}, XML \cite{XMLlib}, ggplot2 \cite{ggplot2}, GGally \cite{GGally}, and gridExtra \cite{gridExtra} for data curation and visualisation.

\section*{Acknowledgments}
J.M.C. is supported by the BBSRC and University of Birmingham via the MIBTP doctoral training scheme (grant number BB/M01116X/1). This project has received funding from the European Research Council (ERC) under the European Union's Horizon 2020 research and innovation programme (grant agreement no. 805046 (EvoConBiO) to I.G.J.). The authors are grateful to Morten Brun and Stein Andreas Bertelsen for useful discussions, and to Stors{\aa}ta for inspiration. We gratefully acknowledge the Imaging Suite (BALM) at the University of Birmingham for support of imaging experiments and thank Alessandro di Maio and Prof. Markus Schwarzl\"{a}nder for advice with design and analysis of imaging experiments.

\bibliographystyle{apalike}
\bibliography{plantmito}

\newpage

\section*{Supplementary Information}

\renewcommand\thefigure{S\arabic{figure}}    
\setcounter{figure}{0}



\begin{figure*}
  \includegraphics[width=\textwidth]{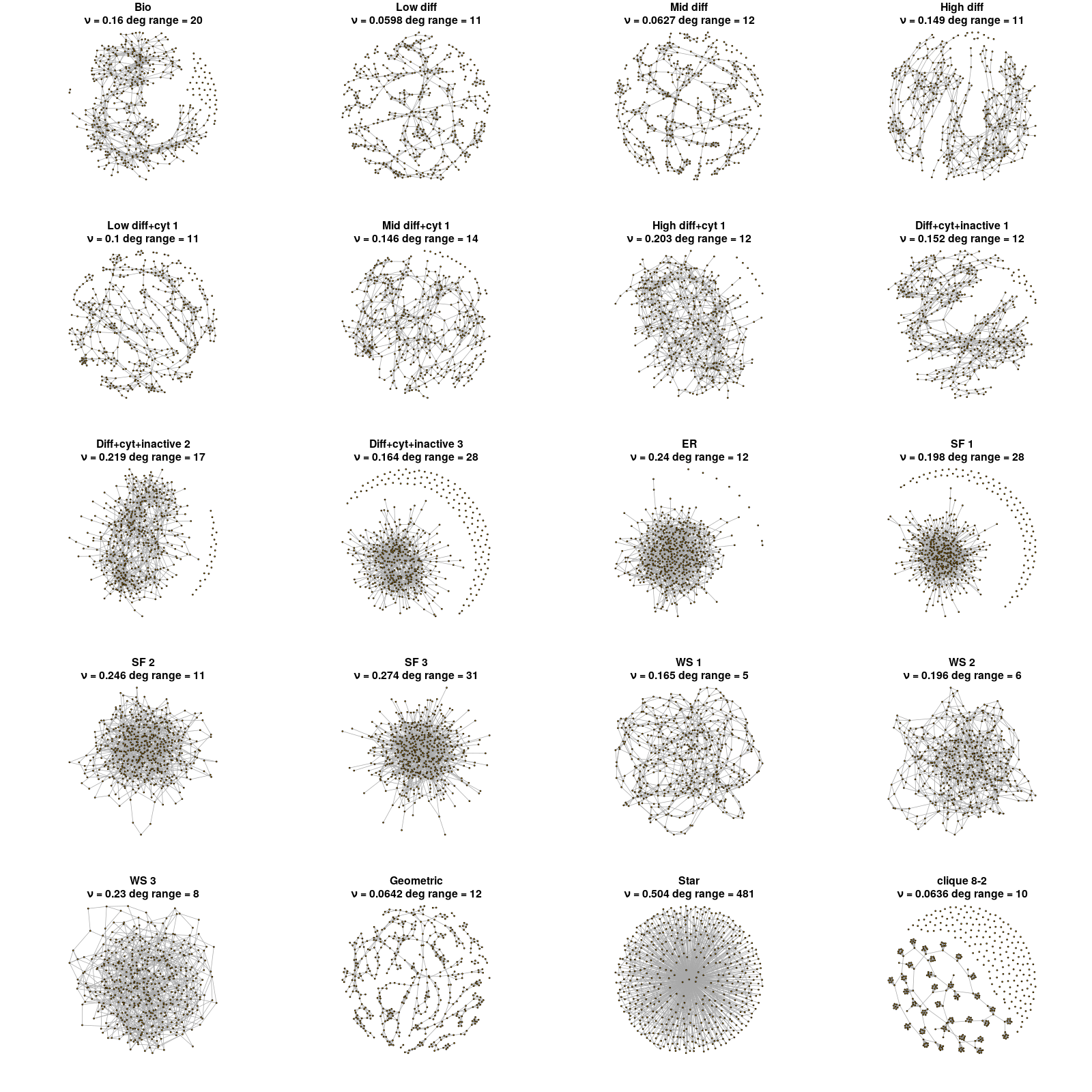}
\caption{\textbf{Comparison of encounter networks from experiment, simulation of mitochondrial dynamics, and general theory.} Visualisations of network structures from the different construction protocols described in Methods, matching (as closely as possible) the statistics of the biological (`Bio') network. One representative `cliquey' network structure is shown; abbreviations are ER (Erd\H{o}s-R\'{e}nyi), SF (scale-free), WS (Watts-Strogatz). Network statistics are $\nu$, global efficiency; and \emph{deg range}, range of degree distribution.}
  \label{figsi-graphs}
\end{figure*}

\begin{figure*}
  \includegraphics[width=\textwidth]{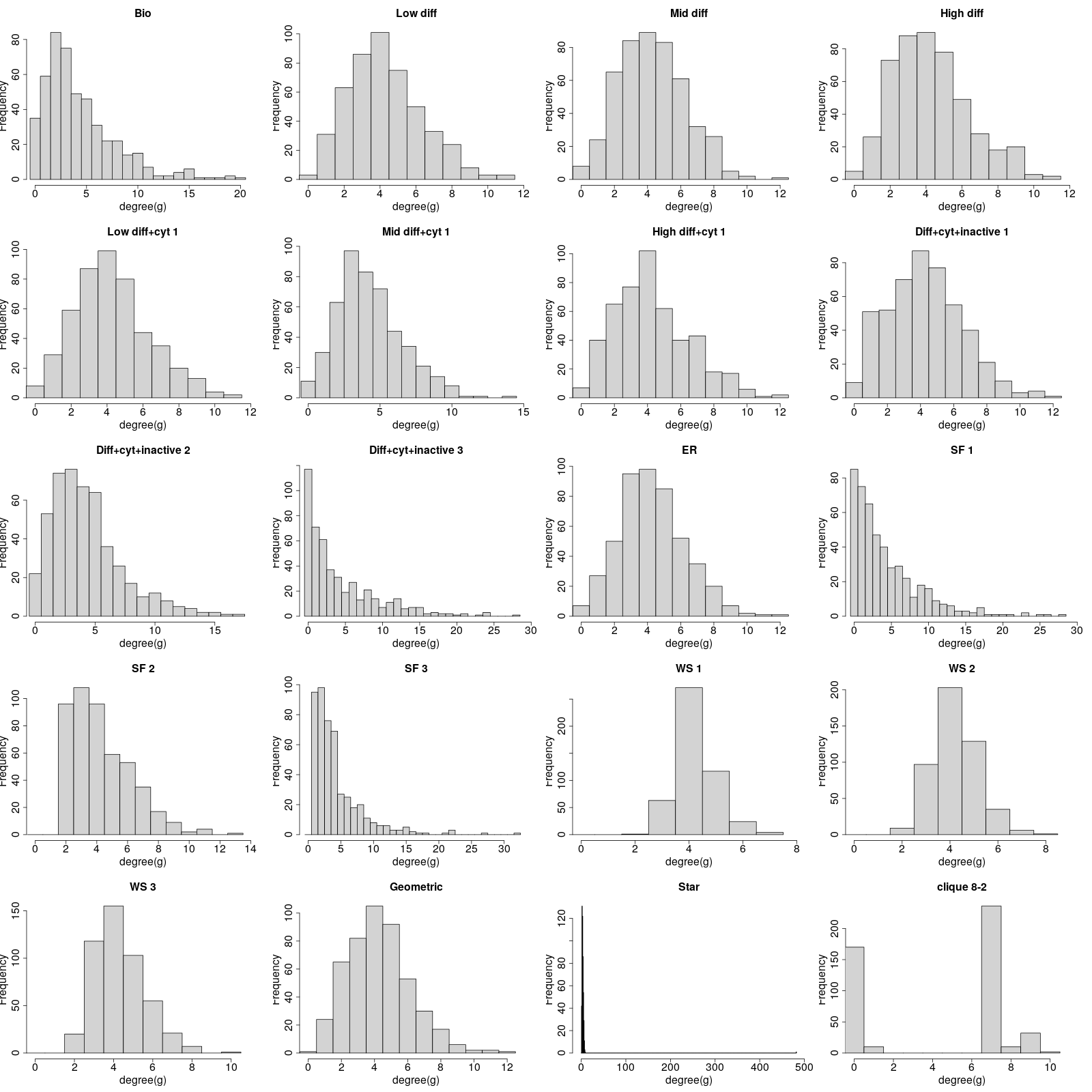}
\caption{\textbf{Degree distributions of encounter networks from experiment, simulation of mitochondrial dynamics, and general theory.} Degree distributions for the ensemble of graphs in Fig. \ref{figsi-graphs}. One representative `cliquey' network structure is shown; abbreviations are ER (Erd\H{o}s-R\'{e}nyi), SF (scale-free), WS (Watts-Strogatz).}
  \label{figsi-degrees}
\end{figure*}

\begin{figure*}
  \includegraphics[width=\textwidth]{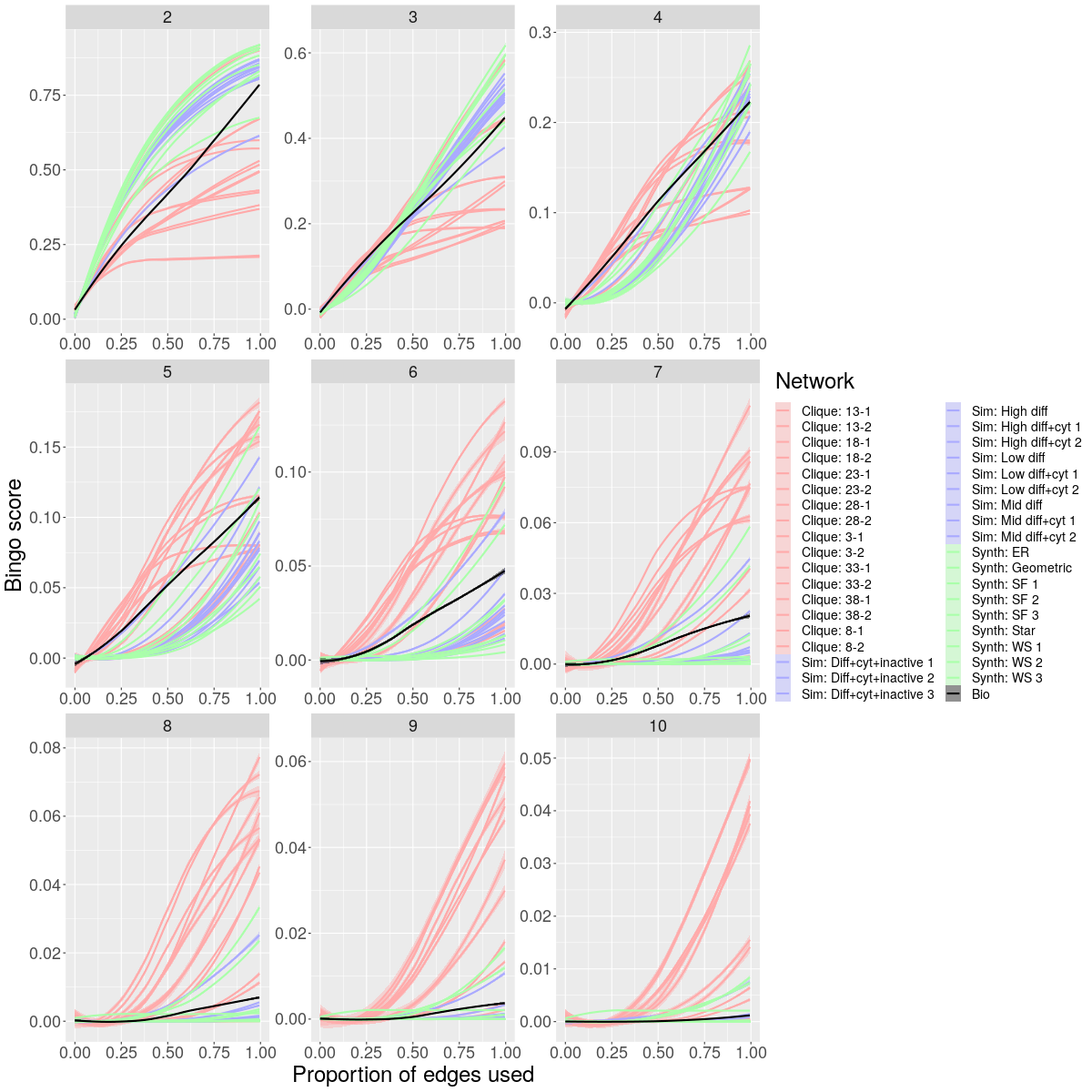}
\caption{\textbf{Bingo dynamics on different networks for different $L$ with $m = 0$.} Behaviour of bingo score $p$ with proportion of edges $q$ used for genetic exchange, arranged for a range of synthetic networks and their biological partner, for different $L$. Traces are coloured by the general class of synthetic network. Traces are LOESS fits to $n = 10$ simulations for each case.}
  \label{figsi-dynamics}
\end{figure*}

\begin{figure*}
  \includegraphics[width=0.7\textwidth]{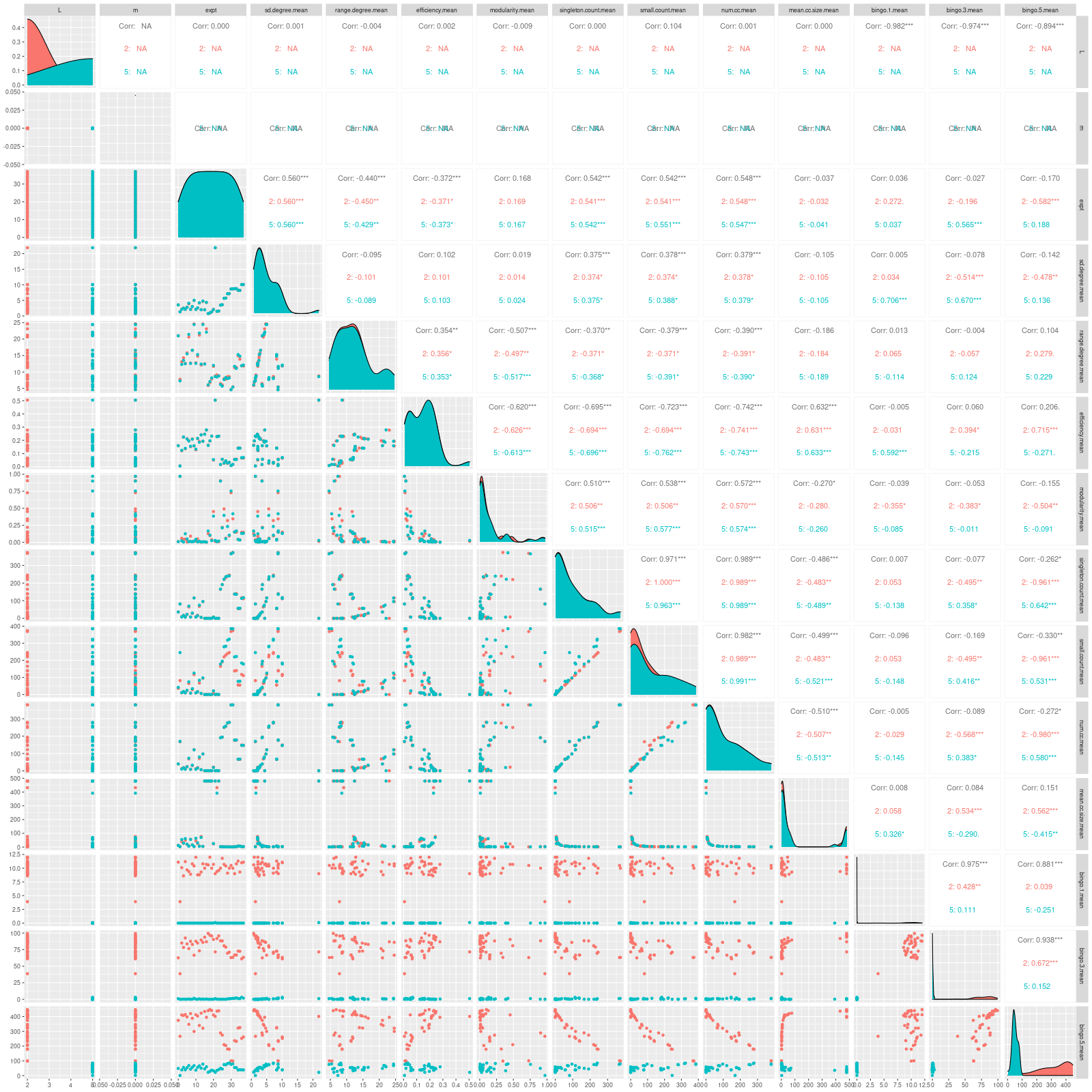}
  \caption{\textbf{Network statistics and bingo performance.} Correlations between network statistics and bingo performance, for $L=2$ (red) and $L=5$ (blue), with scatter plots under the diagonal, Pearson coefficients above the diagonal (${}^{*}, p < 0.05$; ${}^{**}, p < 0.01$; ${}^{***}, p < 0.001$), and histograms of the statistic on the diagonal. Each point is a mean value for a different class of network, taken over 10 generated instances. Labels: \emph{sd.degree} and \emph{range.degree}, degree distribution standard deviation and range; \emph{efficiency}, global network efficiency; \emph{modularity}, network modularity measured using the walktrap algorithm \cite{pons2006computing}; \emph{singleton.count}, number of degree-zero nodes; \emph{small.count}, number of components with size $<L$; \emph{num.cc} and \emph{mean.cc.size}, number and mean size of connected components; \emph{bingo.1/3/5}, bingo score when proportion 0.01/0.1/1 of edges are used for genetic exchange. Although some statistics correlate with bingo performance for a given $L$, little correlation across $L$ values is visible.}
  \label{figsi-correlations}
\end{figure*}

\begin{figure*}
  \includegraphics[width=\textwidth]{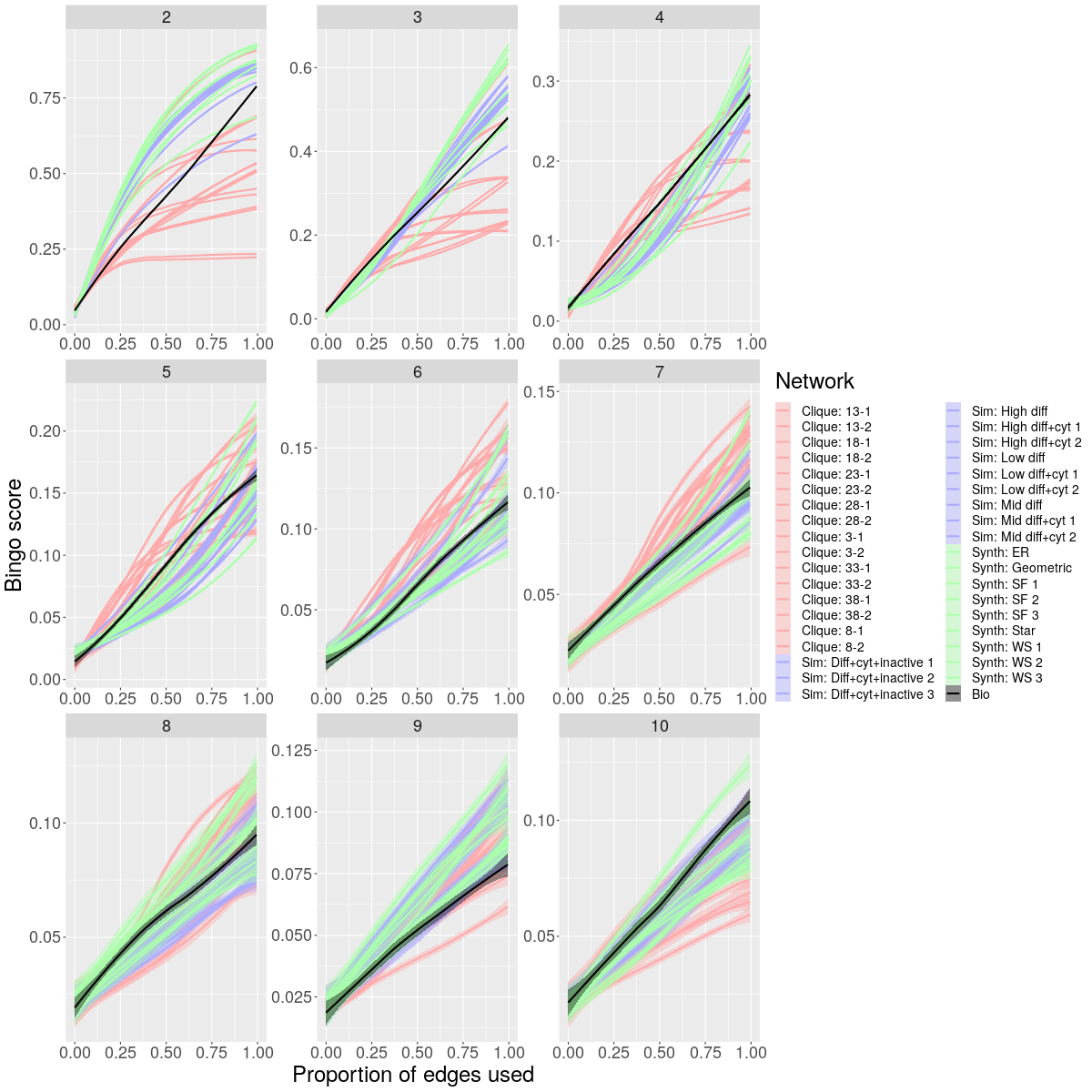}
\caption{\textbf{Bingo dynamics on different networks for different $L$ with $m = 0.02$.} Analogous to Fig. \ref{figsi-dynamics}, behaviour of bingo score $p$ with proportion of edges $q$ used for genetic exchange, arranged for a range of synthetic networks and their biological partner, for different $L$ and a master circle proportion of $m = 0.02$. Traces are coloured by the general class of synthetic network. Traces are LOESS fits to $n = 10$ simulations for each case.}
  \label{figsi-dynamicsm}
\end{figure*}

\begin{figure*}
  \includegraphics[width=0.5\textwidth]{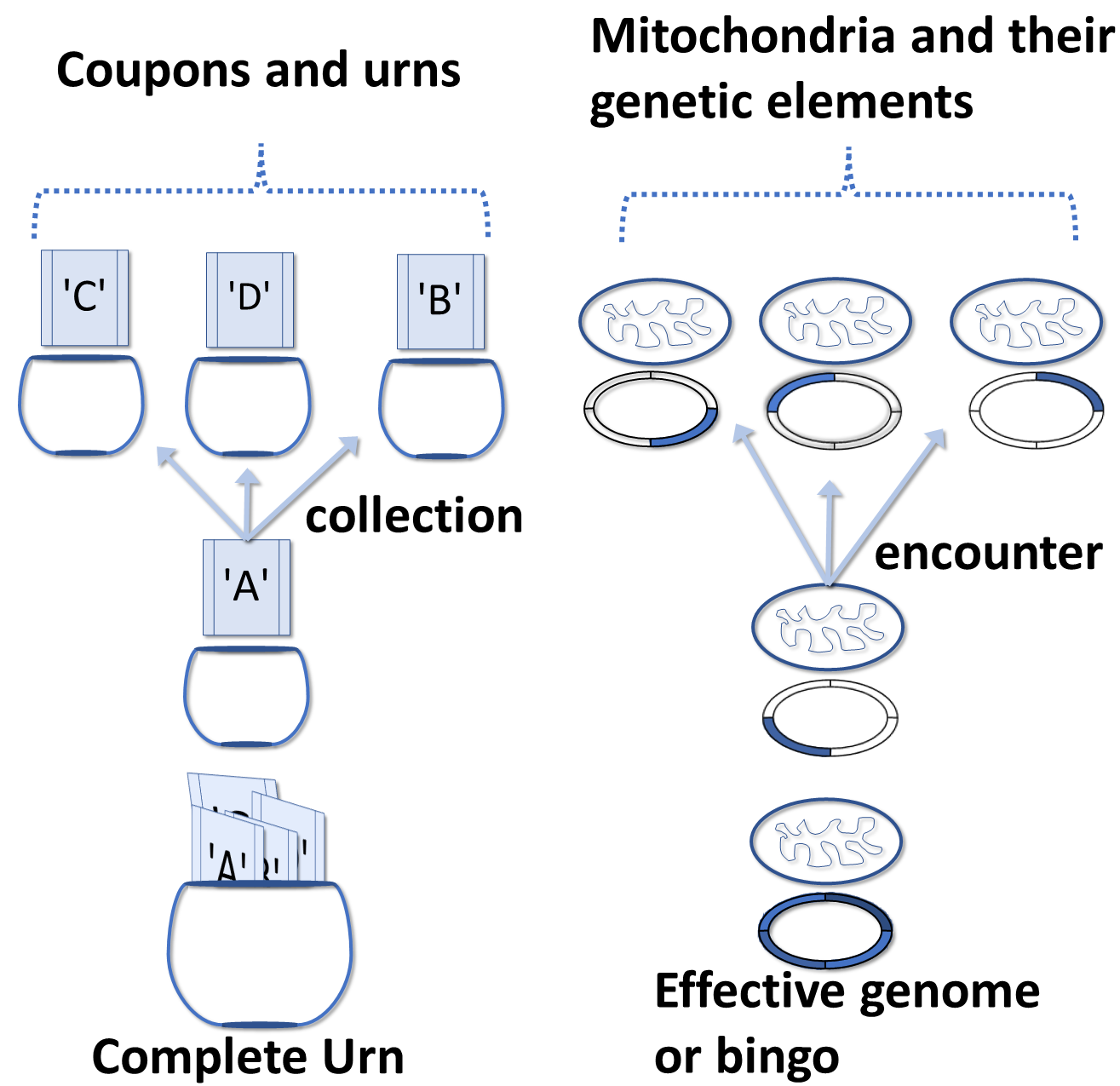}
\caption{\textbf{The analogy between the CCP and bingo.} It is shown and explained the shared terminology between the two concepts and how the coupon collection corresponds to the the assembly of effective genome through encounters with partial genome elements.}
  \label{figsi-bingoAnalogy}
\end{figure*}

\begin{figure*}
  \includegraphics[width=.49\textwidth]{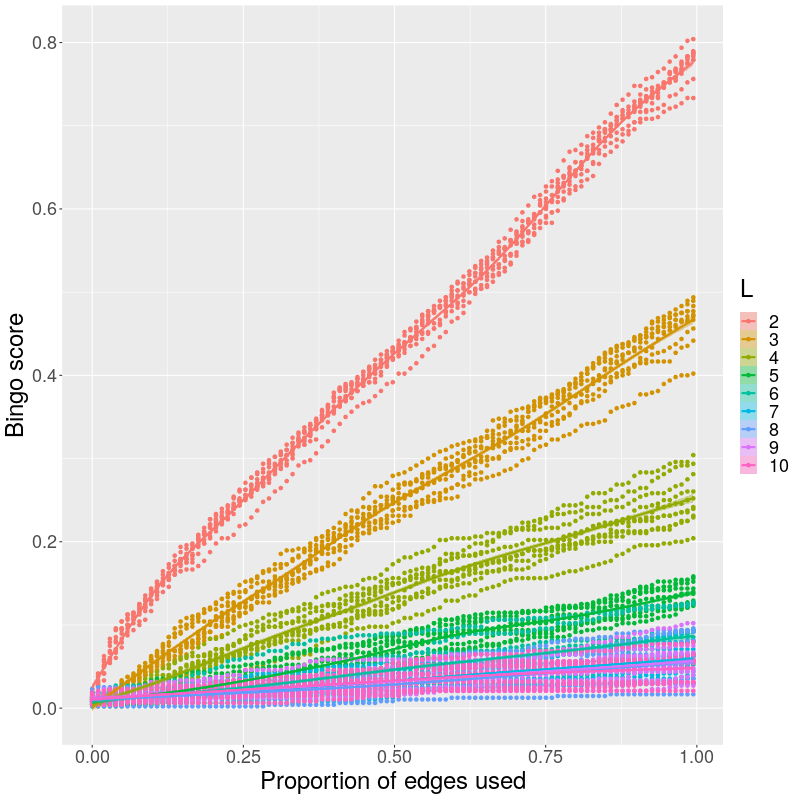}  \includegraphics[width=.49\textwidth]{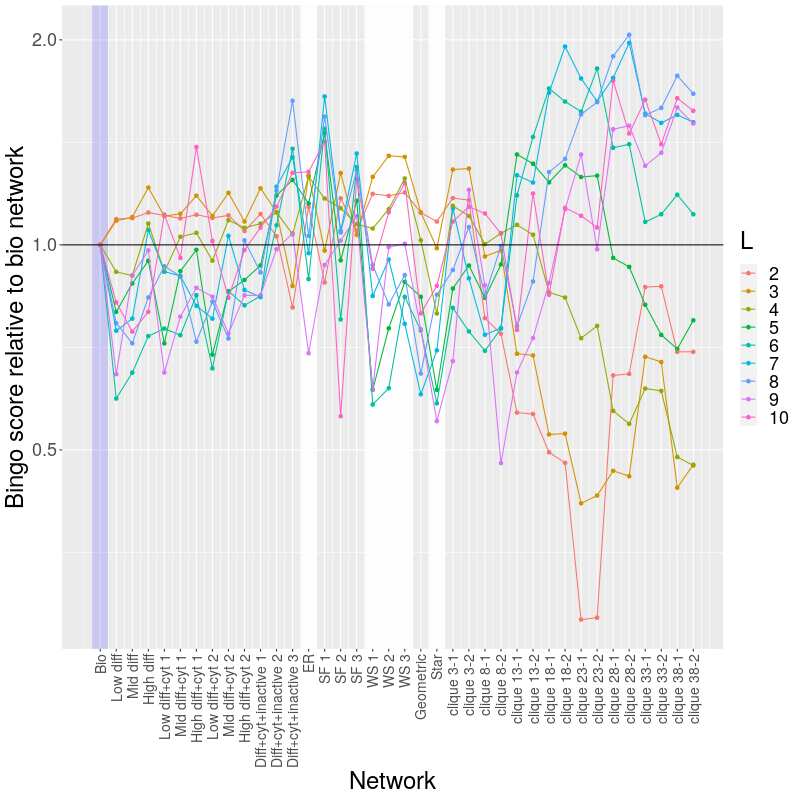}
      \includegraphics[width=.49\textwidth]{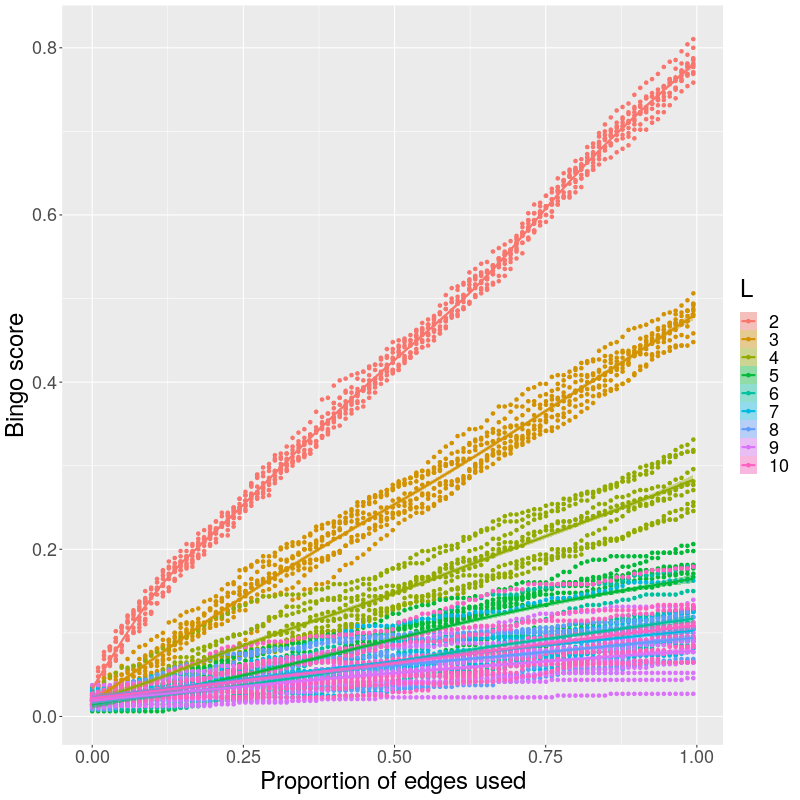}  \includegraphics[width=.49\textwidth]{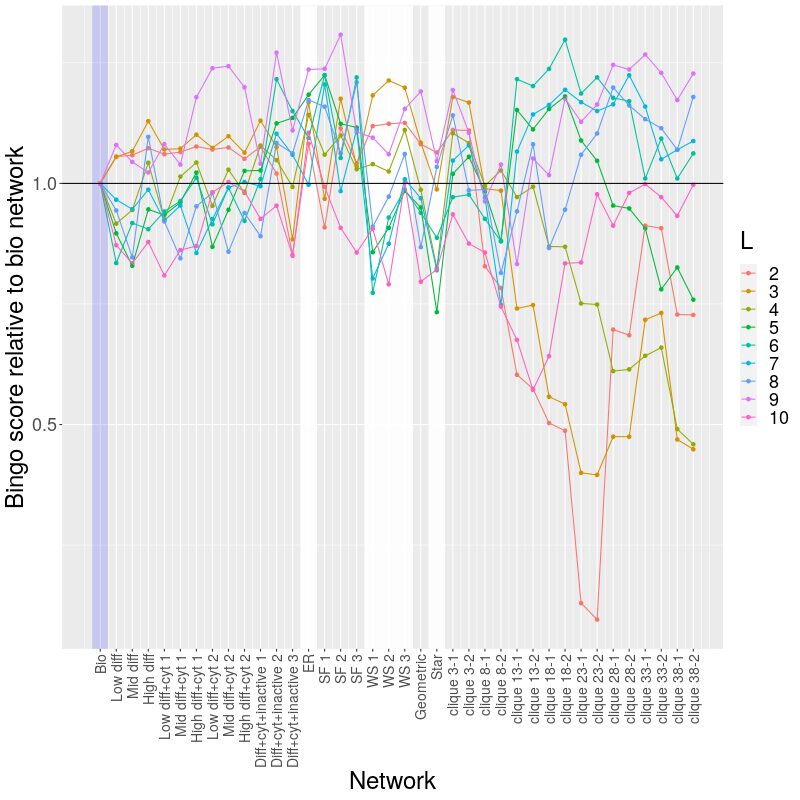}
\caption{\textbf{Effective genome emergence in the presence of master circles.} Bingo score with proportion of edges used for genetic exchanges (analogous to Fig. \ref{fig1}) and final bingo score compared to synthetic networks with matched statistics (analogous to Fig. \ref{fig2a}), for different $L$. (top) a proportion $m = 0.01$ of genetic elements are `master circles', providing all genetic elements at once; (bottom) $m = 0.02$. Bingo performance is generally increased and more comparable across networks than in the absence of master circles.}
  \label{figsi-bingom}
\end{figure*}

\begin{figure*}
  \includegraphics[width=.3\textwidth]{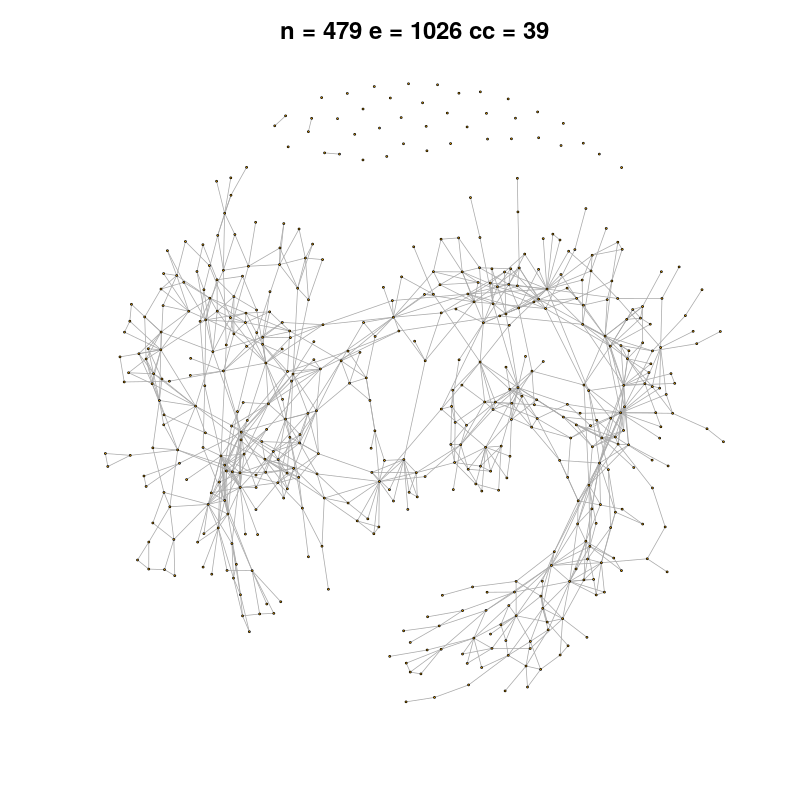}  \includegraphics[width=.3\textwidth]{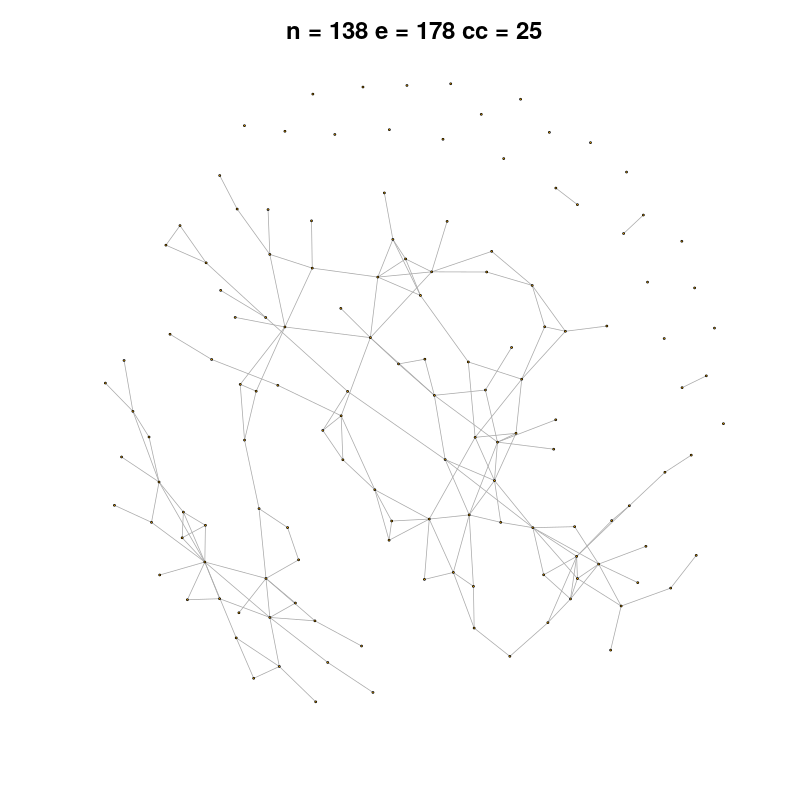}  \includegraphics[width=.3\textwidth]{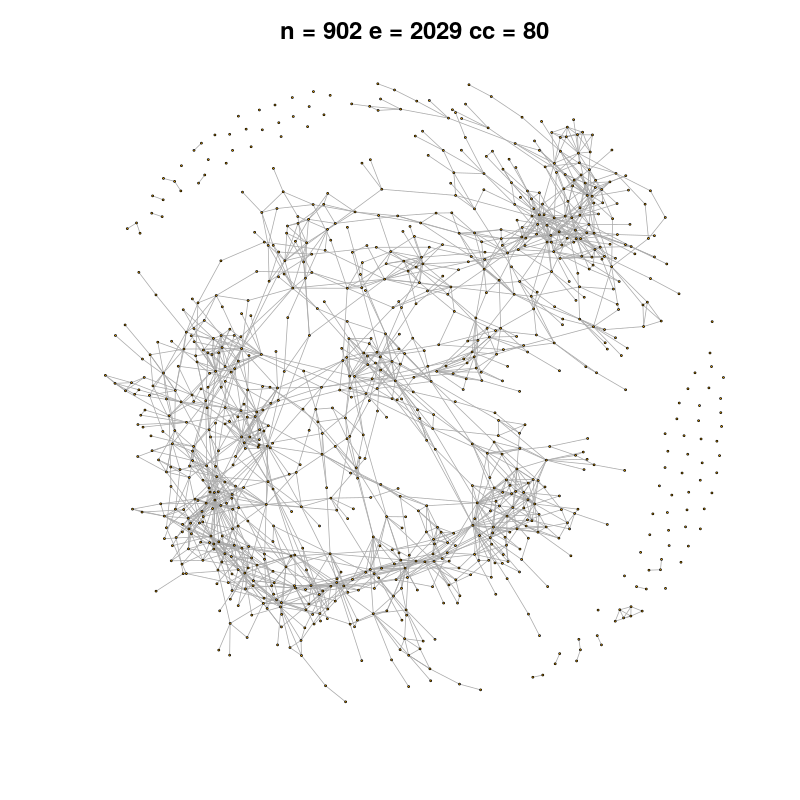} \\
  \includegraphics[width=.3\textwidth]{orig-1.xml-amlist.csv-0-0-results-overall.txt.results-m-0.png}   \includegraphics[width=.3\textwidth]{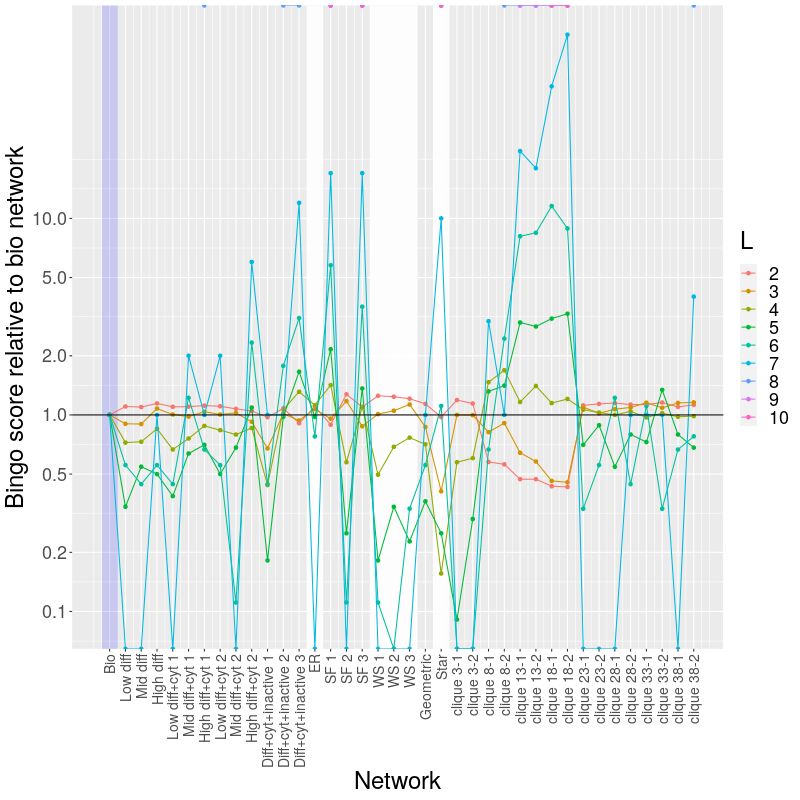}   \includegraphics[width=.3\textwidth]{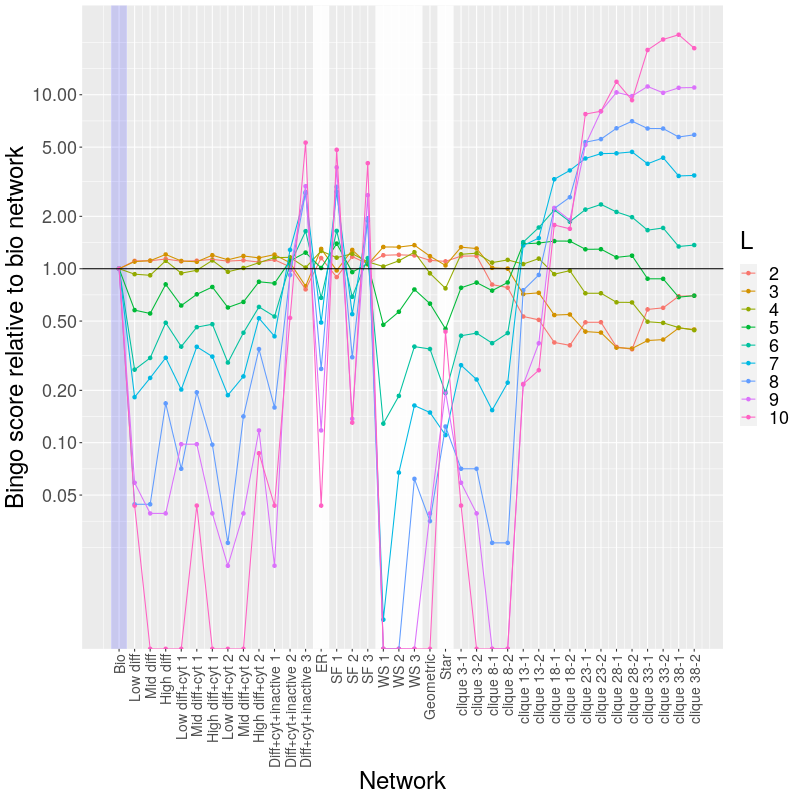}
  \includegraphics[width=.3\textwidth]{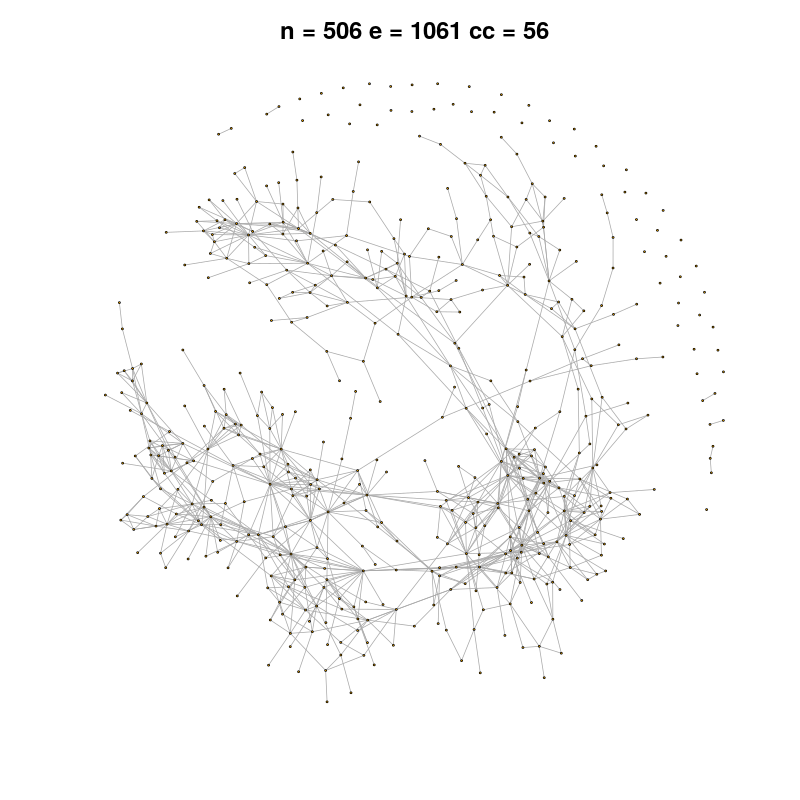}  \includegraphics[width=.3\textwidth]{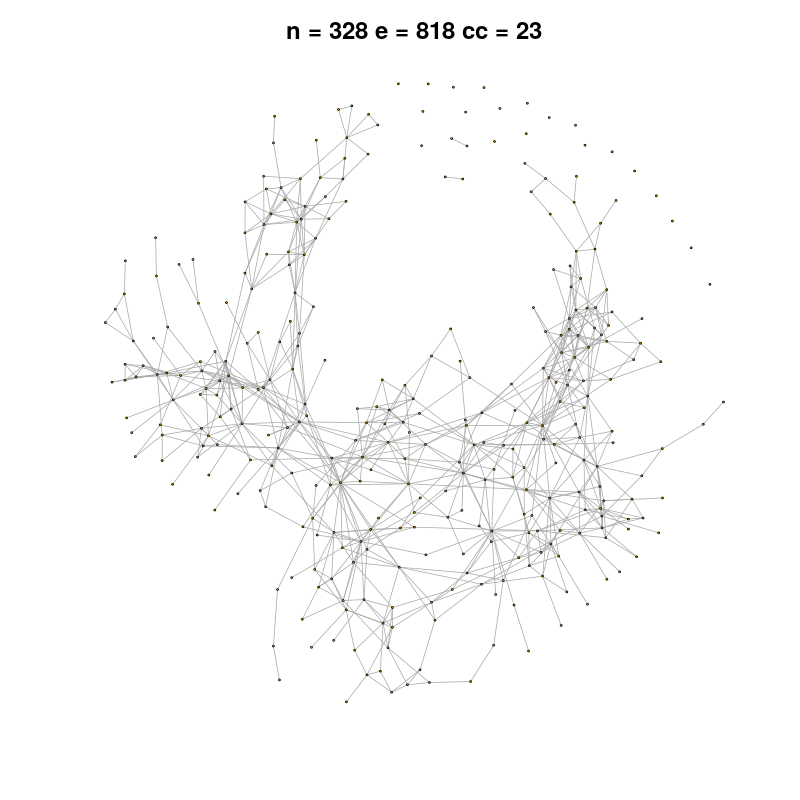}  \includegraphics[width=.3\textwidth]{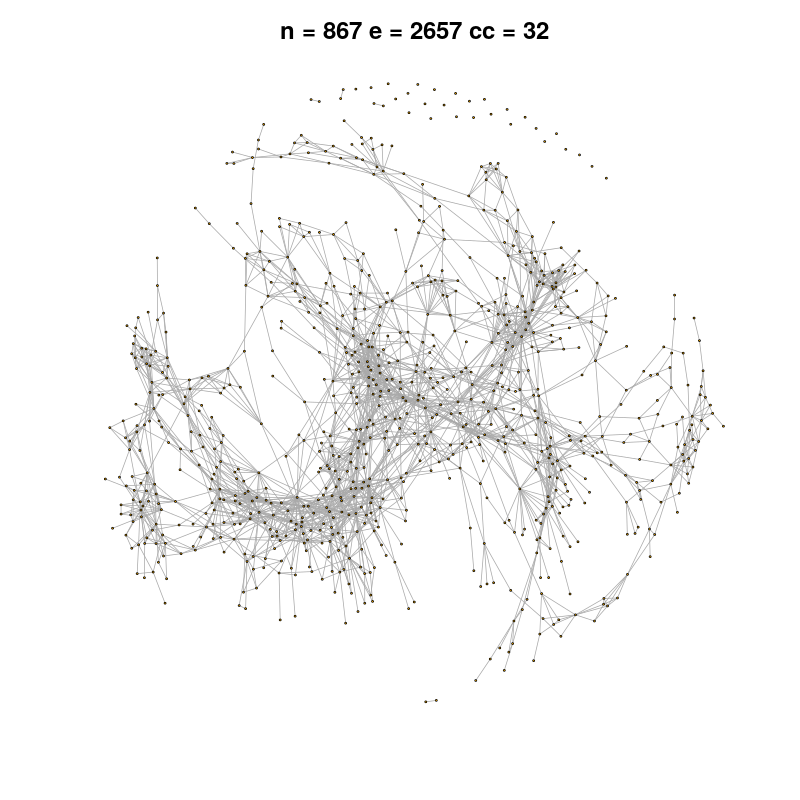} \\
  \includegraphics[width=.3\textwidth]{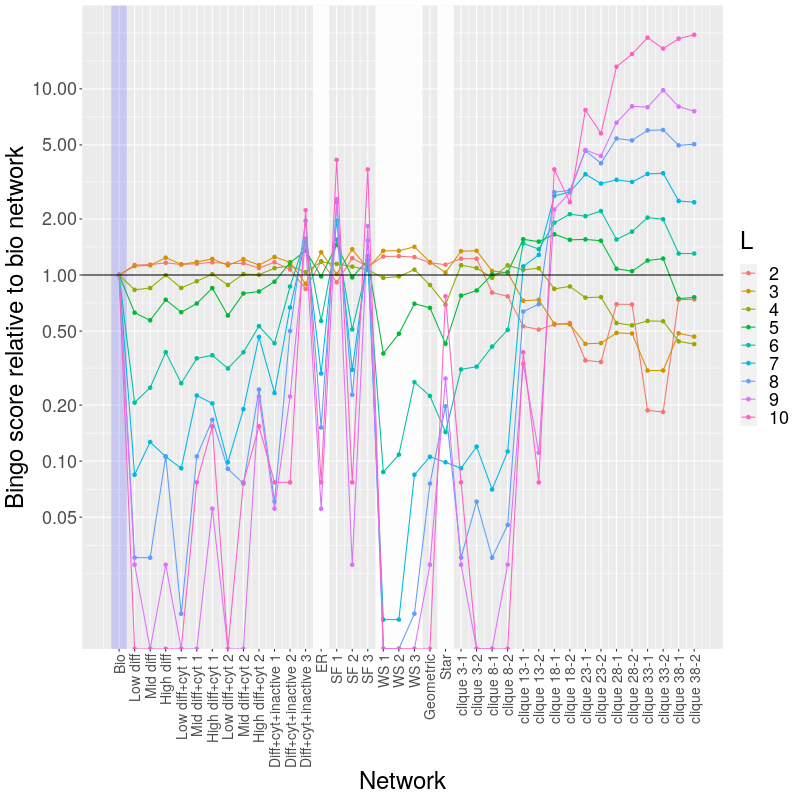}   \includegraphics[width=.3\textwidth]{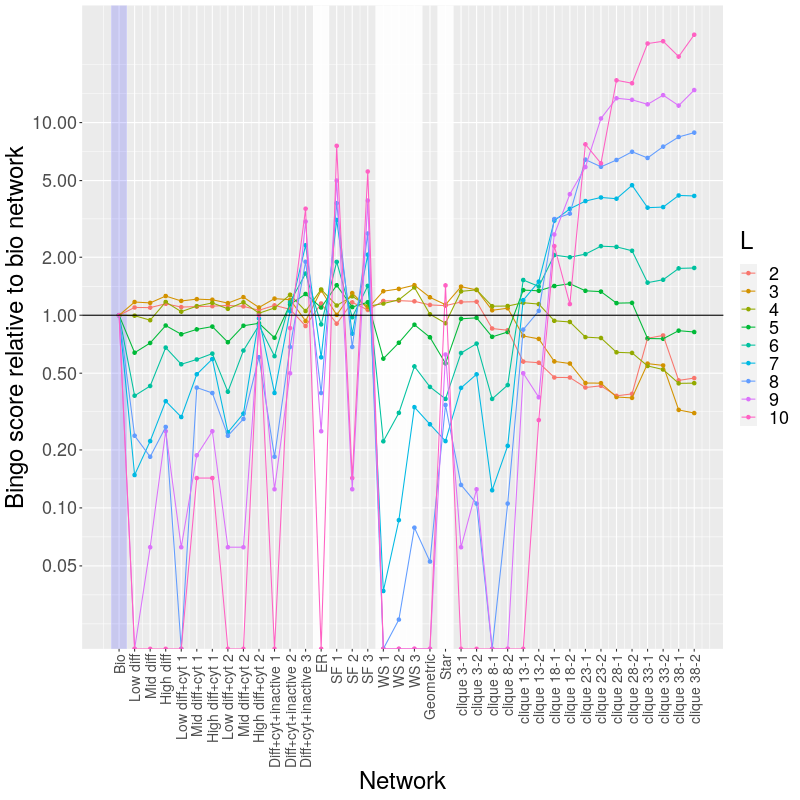}   \includegraphics[width=.3\textwidth]{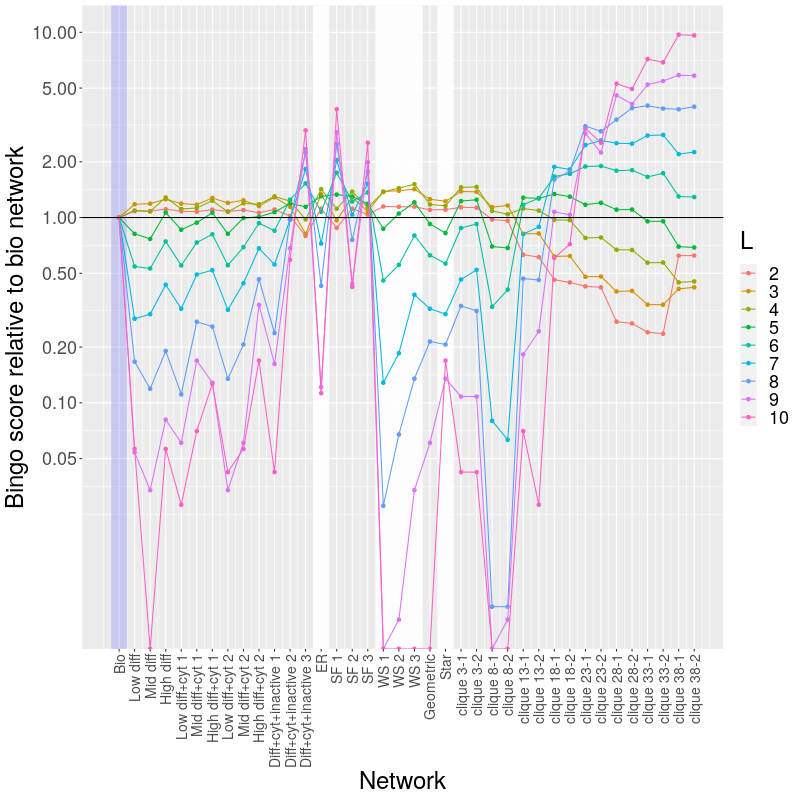}

  \caption{\textbf{Comparison of behaviour across different cells.} Examples of encounter network visualisations ($n$ nodes, $e$ edges, $cc$ connected components) and profiles of biological \emph{versus} synthetic partner bingo performance (analogous to Fig. \ref{fig2a}) for different single cell observations. All but the top centre cell show very comparable trends. The top centre was unusually small, limiting the size of the mitochondrial population and hence the scale of the encounter network. Correspondingly, the bingo performance for both biological and synthetic partner networks is diminished, especially for high $L$, but the relative performance trends remain comparable.}
  \label{figsi-differentcells}
\end{figure*}

\begin{figure*}
  \includegraphics[width=.33\textwidth]{orig-1.xml-amlist.csv-0-0-results-overall.txt.results-m-0.png}  \includegraphics[width=.33\textwidth]{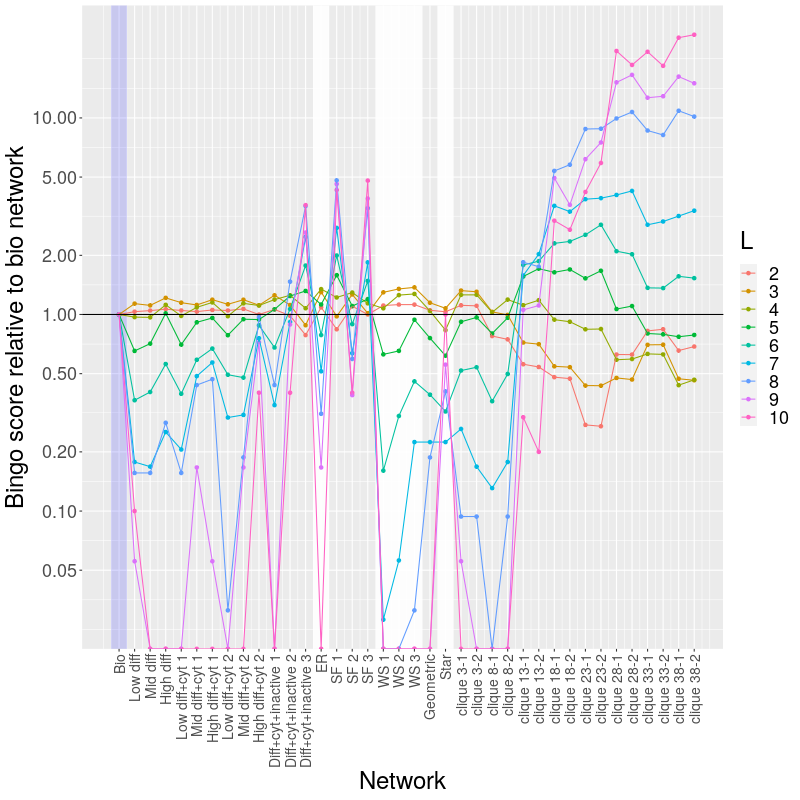}  \includegraphics[width=.33\textwidth]{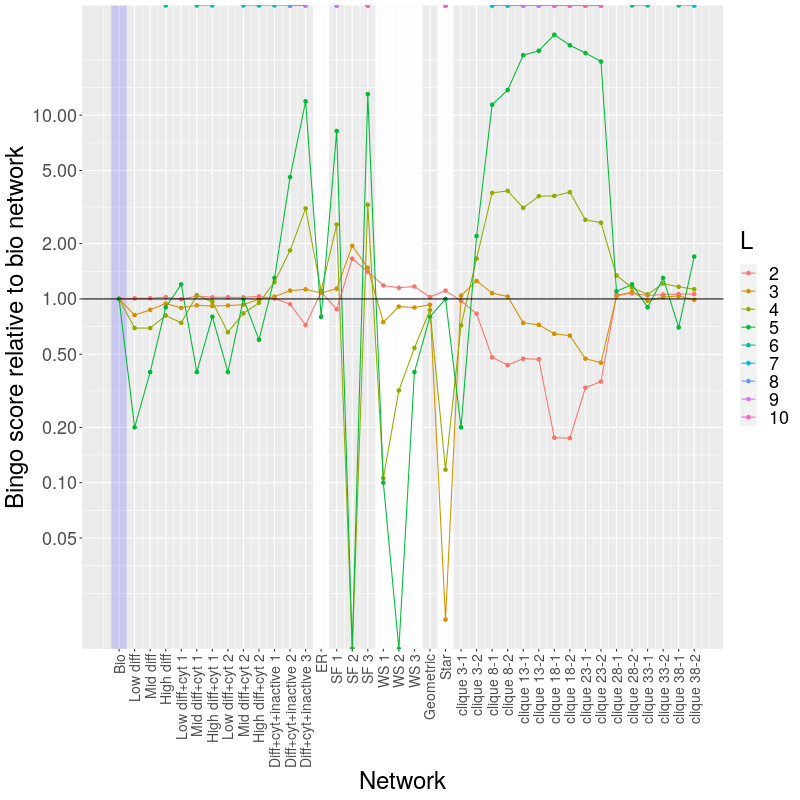}
  \caption{\textbf{Comparison of behaviour in different circumstances.} Analogous to Fig. \ref{fig2a}, bingo performance for a number of changes to the experimental setup. Left, normal; centre, singletons removed from biological encounter network; right, biological trajectories pruned to a maximum length of ten frames (23 seconds). }
  \label{figsi-differentexpts}
\end{figure*}

\end{document}